\documentclass[12pt]{article}
\pdfoutput=1

\usepackage[utf8]{inputenc}

\usepackage{draft}
\usepackage{putex}
 
 \usepackage[table]{xcolor}
 \usepackage{dsfont}
\usepackage[all]{xy}

\usepackage{stackrel,amssymb}

\usepackage{graphicx}
\usepackage{caption}
\usepackage{amsmath,bm}
\usepackage{array}
\usepackage{subcaption}
\usepackage{epstopdf}
\usepackage{enumerate}
\usepackage{cite}
\usepackage{tensor}
\usepackage{slashed}
\usepackage[utf8]{inputenc}
\usepackage{rotating}
\usepackage{bigfoot}
\usepackage[
      colorlinks=true,
      linkcolor=blue,
      urlcolor=blue,
      filecolor=black,
      citecolor=red,
      ]{hyperref}

\usepackage{tikz-cd}

\def\XXint#1#2#3{{\setbox0=\hbox{$#1{#2#3}{\int}$ }
		\vcenter{\hbox{$#2#3$ }}\kern-.6\wd0}}

  \usepackage{adjustbox}
\usepackage{multirow}
\usepackage{tikz-cd}

\numberwithin{equation}{section}

\newcommand{\cI}{{\cal I}}

\newtheorem{theorem}{Theorem}

\def\<{\langle}
\def\>{\rangle}

\def\pa{\partial}

\def\ep{\epsilon}

\def\id{\mathds{1}}

\newcommand{\leftrarrows}{\mathrel{\raise.75ex\hbox{\oalign{%
				$\scriptstyle\leftarrow$\cr
				\vrule width0pt height.5ex$\hfil\scriptstyle\relbar$\cr}}}}
\newcommand{\lrightarrows}{\mathrel{\raise.75ex\hbox{\oalign{%
				$\scriptstyle\relbar$\hfil\cr
				$\scriptstyle\vrule width0pt height.5ex\smash\rightarrow$\cr}}}}
\newcommand{\Rrelbar}{\mathrel{\raise.75ex\hbox{\oalign{%
				$\scriptstyle\relbar$\cr
				\vrule width0pt height.5ex$\scriptstyle\relbar$}}}}

\makeatletter
\def\leftrightarrowsfill@{\arrowfill@\leftrarrows\Rrelbar\lrightarrows}
\newcommand{\xleftrightarrows}[2][]{\ext@arrow 3399\leftrightarrowsfill@{#1}{#2}}
\makeatother

\begin{document}

\title{Gauging Non-Invertible Symmetries:
Topological Interfaces and Generalized Orbifold Groupoid in 2d QFT
}

\authors{Oleksandr Diatlyk, Conghuan Luo, Yifan Wang, and Quinten Weller}

\institution{NYU}{Center for Cosmology and Particle Physics, New York University, New York, NY 10003, USA}

\abstract{
Gauging is a powerful operation on symmetries in quantum field theory (QFT), as it connects distinct theories and also reveals hidden structures in a given theory.
We initiate a systematic investigation of gauging discrete generalized symmetries in two-dimensional QFT. Such symmetries are described by topological defect lines (TDLs) which obey fusion rules that are non-invertible in general. Despite this seemingly exotic feature, all well-known properties in gauging invertible symmetries carry over to this general setting, which greatly enhances both the scope and the power of gauging. This is established by formulating generalized gauging in terms of topological interfaces between QFTs, which explains the physical picture for the mathematical concept of algebra objects and associated module categories over fusion categories that encapsulate the algebraic properties of generalized symmetries and their gaugings. This perspective also provides simple physical derivations of well-known mathematical theorems in category theory from basic axiomatic properties of QFT in the presence of such interfaces. We discuss a bootstrap-type analysis to classify such topological interfaces and thus the possible generalized gaugings and demonstrate the procedure in concrete examples of fusion categories. Moreover we present a number of examples to illustrate generalized gauging and its properties in concrete conformal field theories (CFTs). In particular, we identify the generalized orbifold groupoid that captures the structure of fusion between topological interfaces (equivalently sequential gaugings) as well as a plethora of new self-dualities in CFTs under generalized gaugings.

}
\date{}

\maketitle

\tableofcontents

\section{Introduction and Summary}

Symmetry is arguably the most powerful principle in physics and a crucial concept ingrained in the foundation of Quantum Field Theory (QFT). 
There is plenty of evidence that QFT dynamics is extremely rich from both experiments and numerical simulations, thanks to a myriad of strong coupling effects. However in a typical parameter regime, a controlled calculation is often not possible and a naive extrapolation can be misleading. This is where symmetry plays an important role. It allows us to bypass the barrier of strong interactions and deduce nontrivial dynamical properties of physical systems, ranging from selection rules on probability amplitudes, constraints on phase transitions, to predictions for the destinies of renormalization group flows. This is possible by virtue of a notion of rigidity associated with symmetries which produces simple mathematical invariants for complicated QFT observables, and such invariants can often be calculated by going to weakly-coupled corners in the parameter space. 
These invariants encode finer features of symmetries, 
such as the 't Hooft anomaly for a symmetry group $G$, 
and a thorough examination of such invariants can greatly strengthen the predictions from symmetries on the QFT dynamics. It is well known that a universal and elegant way to probe such invariants of symmetry is to couple the QFT to background gauge fields; the 't Hooft anomalies are then obstructions to gauging, which corresponds to promoting these gauge fields to become dynamical while satisfying the basic axioms of QFT. When the symmetry $G$ of a QFT $\cT$ is anomaly free, gauging $G$ produces another healthy QFT $\cT/G$, and many known QFTs are generated this way. Various different parts of the QFT theory space are then connected by gauging. This can be used to shed light on the properties of one QFT from another, which may be obscure without this extra perspective.

The past decade has witnessed a major paradigm shift in how to think about symmetries, leading to the notion of generalized symmetries represented by topological defect operators $\cD(\Sigma_{(p)})$ in a QFT defined over a $p$-dimensional closed submanifold $\Sigma_{(p)}$ of the spacetime \cite{Gaiotto:2014kfa} (see also \cite{McGreevy:2022oyu,Cordova:2022ruw,Schafer-Nameki:2023jdn,Brennan:2023mmt,Bhardwaj:2023kri,Shao:2023gho} for recent lecture notes and reviews). In this language, ordinary symmetries correspond to a family of topological defects $\cD_g(\Sigma_{(d-1)})$ defined on codimension-one submanifolds and labeled by group elements $g\in G$.
There are two immediate directions to generalize the above: one could consider topological defects of higher codimensions or consider topological defects no longer labeled by group elements. The former has led to the notion of higher form symmetries whereas the latter has ushered in the non-invertible symmetries. 
Given the importance of gauging to both detecting anomalies 
and relating potentially different QFTs, it is imperative to understand how to implement gauging for these generalized symmetries. This question is particularly intriguing for non-invertible symmetries, as
the composition of corresponding symmetry transformations now carries a fusion algebra structure generalizing that of a group and thus requires a new mathematical language to describe the corresponding gauge fields and a generalized notion of anomaly to describe potential obstructions. 

Here we focus on the physics of gauging non-invertible symmetries (also known as generalized gauging) in $d=2$ spacetime dimensions. 
In this case, the generators of the symmetry are topological defect lines (TDLs), which we will denote in general by $\cL$ \cite{Buican:2017rxc,Bhardwaj:2017xup,Chang:2018iay}. 
The relevant mathematical language here is well established and given by the theory of fusion categories \cite{etingof2016tensor} (see also the Appendix of \cite{Kitaev:2005hzj} for a concise review). Consequently these symmetries are also sometimes referred to as fusion category symmetries. Each fusion category $\cC$ has a two-layer structure: 
the top layer encodes the TDLs $\cL_i$ as objects in a fusion algebra (i.e. operator product between the TDLs) via the Grothendieck ring $K(\cC)$ of $\cC$; the second layer contains the F-symbols which package finer invariants of the symmetry $\cC$ akin to the 't Hooft anomaly for ordinary finite group symmetry $G$. Indeed, the latter becomes a special case of a fusion category symmetry with $\cC={\rm \bf Vec}_G^\omega$ where $\omega \in H^3(G,U(1))$ characterizes the anomaly, also known as a pointed fusion category. 
Large families of TDLs
have been identified in numerous $d=2$ QFTs and non-invertible symmetries have proven to be at least as ubiquitous as ordinary symmetries \cite{Verlinde:1988sn,Moore:1988qv,Oshikawa:1996dj,Oshikawa:1996ww,Petkova:2000ip,Frohlich:2004ef,Fuchs:2007tx,Frohlich:2006ch,Frohlich:2009gb,Drukker:2010jp,Davydov:2010rm,Brunner:2014lua,Aasen:2016dop,Bhardwaj:2017xup,Buican:2017rxc,Chang:2018iay,Ji:2019ugf,Lin:2019hks,Thorngren:2019iar,Komargodski:2020mxz,Aasen:2020jwb,Chang:2020imq,Huang:2021ytb,Thorngren:2021yso,Huang:2021zvu,Vanhove:2021zop,Burbano:2021loy,Lu:2022ver,Chang:2022hud,Rayhaun:2023pgc,Bashmakov:2023kwo,Haghighat:2023sax,
Duan:2023ykn,Chen:2023jht,Nagoya:2023zky}. 
Their power in constraining QFT dynamics is illustrated in many examples where ordinary symmetries are insufficient \cite{Chang:2018iay,Lin:2019kpn,Thorngren:2019iar,Komargodski:2020mxz,Gaiotto:2020iye,Thorngren:2021yso,Lin:2021udi,Huang:2021zvu,Kaidi:2022cpf,Lin:2023uvm,Zhang:2023wlu,Antinucci:2023ezl,Bhardwaj:2023idu,Bhardwaj:2023fca,Kaidi:2023maf,Sun:2023xxv,Choi:2023xjw}. In particular,
a generalized notion of 't Hooft anomalies for these non-invertible symmetries was put forward in \cite{Thorngren:2019iar} motivated by the requirement that an anomaly should forbid a trivially gapped symmetric ground state, give an efficient way to diagnose the fate of renormalization group flows, and constrain the phase diagram at large distance. As aforementioned, this is thanks to the rigidity of symmetry invariants. In the case of non-invertible symmetries, this rigidity manifests itself as algebraic equations the F-symbols must satisfy and these equations only have finitely many solutions, which is a result know as Ocneanu rigidity in category theory \cite{etingof2016tensor}.
The same rigidity also implies that not all fusion algebras realize full fledged categories which may make one worry about TDLs with spurious fusion rules. Nevertheless, it follows from the basic
axioms of locality and unitarity in QFT together with the  fundamental topological property  that such algebraic constraints are automatically satisfied \cite{Chang:2018iay}, and thus QFT provides a gigantic factory for fusion categories, most of which are beyond the existing classifications in mathematics (see for example \cite{etingof2003classification,rowell2009classification,Jordan_2009,ostrik2013pivotal,larson2014pseudounitary,bruillard2018classification,ediemichell2020classifying} for previous attempts).

\begin{figure}[!htb]
    \centering
   \includegraphics[scale=0.2]{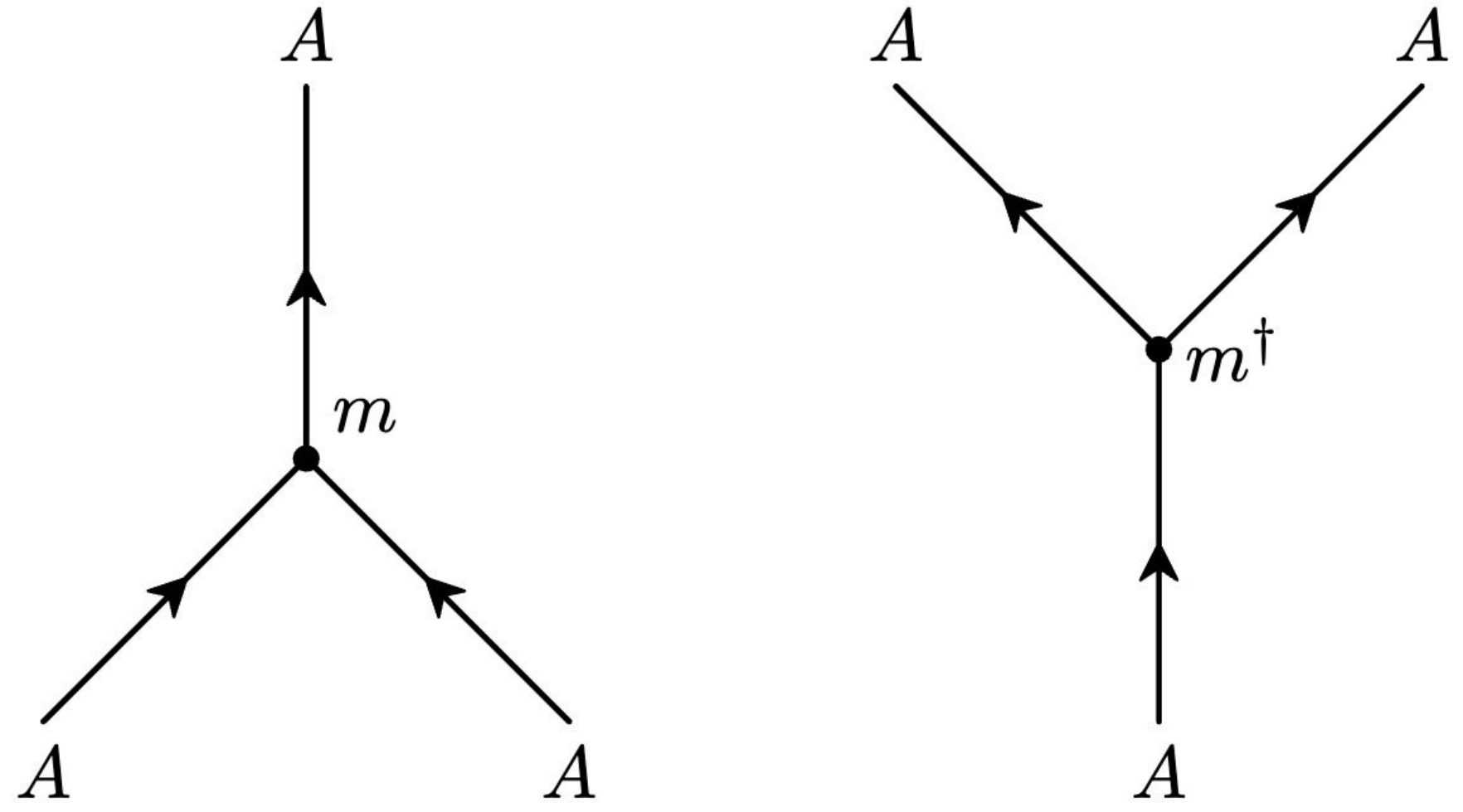}
    \caption{Trivalent topological junction $m$ for the algebra object $A$ which specifies the multiplication morphism and $m^\dagger$ for the comultiplication morphism.}
    \label{fig:multiplication}
\end{figure}

Intuitively, gauging a non-invertible symmetry $\cC$ amounts to decorating the QFT observables with the corresponding TDLs in $\cC$, which generalizes the gauge field for discrete group symmetries (see e.g. Figure~\ref{fig:torus}), and then summing over the TDL configurations created by their topological junctions, subject to consistency conditions that the resulting observables obey basic axioms of QFT such as unitarity and locality. In particular, these consistency conditions enforce gauge invariance that manifests itself through invariance under topological changes of the TDL configuration (Pachner moves in the dual triangulation). For a symmetry represented by a discrete $G$, this amounts to a trivialization of the anomaly $\omega \in H^3(G,U(1))$, i.e. the symmetry has to be non-anomalous. In the more general context of non-invertible symmetries, the criterion differs in a subtle way. Unless one attempts to gauge the whole fusion category, even an anomalous fusion category may admit consistent ways of gauging a subset of the non-invertible TDLs.\footnote{In \cite{Choi:2023xjw}, motivated by the consideration of symmetric boundary conditions, two distinct notions of anomaly-freeness were introduced for non-invertible symmetries described by a fusion category $\cC$. A fusion category $\cC$ is called \textit{strongly} anomaly-free if $\cC$ admits a fiber functor, and it is \textit{weakly} anomaly-free if there exists an algebra object $A\in \cC$ such that every simple TDL appears in $A$. Here in the main text, by anomaly-free we mean the \textit{strongly} anomaly-free condition of \cite{Choi:2023xjw}. 
}
Heuristically this is possible because non-invertible TDLs tend to admit more topological junctions and thus more possibilities for the decorated observables to be physically consistent. Indeed, these consistency conditions (see Figure~\ref{fig:algebraconditions}) pick out special objects in the relevant fusion category, whose full algebraic structure is that of a symmetric separable Frobenius algebra $(A,m)$ with $A\in \cC$ given by a direct sum of a collection of TDLs and $m$ is the so-called multiplication morphism that specifies a particular trivalent topological junction among $A$ (see Figure~\ref{fig:multiplication}) \cite{Fuchs:2002cm}. Gauging $A$ for a $\cC$-symmetry QFT $\cT$ then comes from decorating observables in $\cT$ with TDL networks formed by $A$ with trivalent junctions $m$, which produces observables in the gauged theory. The resulting gauged theory is denoted as $\cT/A$.

It is natural to ask if general properties in gauging ordinary discrete group symmetries continue to hold with non-invertible symmetries. The former is extensively studied in the $d=2$ CFT where it is known as orbifolding \cite{francesco2012conformal}. In particular, orbifolding by a discrete group $G$ maps one CFT $\cT$ to another $\cT/G$ with the same central charge $c$, and relates the local operator spectrum as well as the defect (boundary and interface) spectrum of the two CFTs in a precise manner. Relatedly, orbifolding by a discrete abelian group $G$ produces a dual symmetry $G^\vee \cong G$, known as the quantum symmetry \cite{Vafa:1989ih}, and gauging again by the dual symmetry one recovers the original theory $\cT/G/G^\vee \cong \cT$. Furthermore, when it happens that the theory is self-dual under gauging, namely $\cT/G \cong \cT$, the $G$ symmetry of $\cT$ is enlarged by a non-invertible TDL known as a duality defect \cite{Thorngren:2021yso,Choi:2021kmx}, which generalizes the Kramers-Wannier duality of the Ising CFT for $G=\mZ_2$ \cite{Frohlich:2004ef,Frohlich:2006ch}. All these properties carry over for gauging non-invertible symmetries. Just like the symmetries themselves, their gauging (for discrete symmetries) can be formulated in a purely algebraic manner,  and there is extensive mathematical literature on this in the language of module categories and bimodule categories over the relevant fusion category $\cC$ \cite{ostrik2001module,MUGER200381,mueger2001subfactors,etingof2009fusion,etingof2016tensor} as well as closely related works in the context of $d=3$ Topological Quantum Field Theory (TQFT) described by the modular tensor category (MTC) $Z(\cC)$ via the Drinfeld (quantum) center (see for example \cite{Frohlich:2003hm,Kong:2007yv,Kitaev_2012,Fuchs:2012dt,kong2014anyon}). However, the details and implications in general QFT have not been fully worked out.

One of the main purposes of this work is to 
provide a physical picture for the categorical concepts involved in generalized gauging. This is achieved by formulating gaugings as topological interfaces in QFT.\footnote{We emphasize that this idea is not new (for examples of this idea in specific settings see \cite{Carqueville:2012dk,Bhardwaj:2017xup,changnotes,Thorngren:2021yso,Choi:2021kmx}). However, we will work in the most general setting of $d=2$ QFTs, and to the best of our knowledge many physical consequences we derive from this perspective have not appeared before.} Similar to the way that topological defects naturally lead to generalized symmetries described by fusion categories as a consequence of basic axioms of QFT (unitarity and locality), 
considerations of QFT in the presence of topological interfaces immediately produce the algebraic structures of module categories. Further mathematical structures are encoded in the fusion of the topological interfaces. In particular the algebra objects arise from the fusion of the corresponding topological interface and its dual, while the fusion between general topological interfaces captures the physics of sequential gauging. The full fusion structure of interfaces is encapsulated by a groupoid which we refer to as the generalized orbifold groupoid generalizing the work of \cite{Gaiotto:2020iye}.
 This physical perspective also motivates a bootstrap type procedure to classify topological interfaces (thus module categories) by exploring consistency conditions from interface fusion. The constraints produced are quite powerful and in the examples we have studied they essentially pin down the module categories up to a few spurious possibilities.\footnote{Because of the correspondence between $\cC$-module categories and $\cC$-symmetric TQFTs in $d=2$ \cite{Thorngren:2019iar,Gaiotto:2020iye} by a slab construction using the 3d TQFT associated with $Z(\cC)$, our results also have implications for the classification of gapped $\cC$-symmetric phases in $d=2$ \cite{DLWW2}.} 
Another main goal here is to provide explicit examples of nontrivial QFTs to illustrate the aforementioned general properties of gauging non-invertible symmetries. 
Moreover we will take advantage the richness in gauging the non-invertible symmetries to find many new TDLs even in familiar CFTs, and we will provide one example here in Section~\ref{sec:SU210TDL}; further examples will be discussed in \cite{DLWW2}.
Finally generalized gauging can relate very different looking renormalization group (RG) flows, and thus we can deduce constraints on families of RG flows from just one member in each family \cite{DLWW2}.

The rest of the paper is organized as follows. In Section~\ref{sec:genprop}, we briefly review non-invertible symmetries, their mathematical description in terms of a fusion category, and describe general properties of non-invertible gauging using the physical language of topological interfaces. In Section~\ref{sec:gengaugingRep}, we discuss explicit examples of fusion category symmetries where we classify all possible gaugings and their corresponding algebraic properties by a bootstrap-type analysis motivated by this physical perspective.
In Section~\ref{sec:CFTrealization}, we employ these algebraic results to understand generalized gauging in several nontrivial CFTs. Along the way 
we give explanations to seemingly miraculous self-dualities under non-invertible gauging, provide physical realizations of the generalized orbifold groupoid for fusion categories, and discover new non-invertible symmetries in familiar CFTs.

{\it Note Added}: As this work was being finalized, and was presented by one of the authors, YW, at \cite{SPOCK}, an independent paper \cite{Choi:2023vgk} appeared on \texttt{arXiv} which has some overlap (especially Section~\ref{sec:Isingsq}) with our results. We have attempted to minimize the overlap in the published version here. The submission of this paper is also coordinated with \cite{PRSVY}, which studies a similar subject.

\section{General Properties of Non-Invertible Symmetries and Their Gauging}
\label{sec:genprop}

\subsection{Topological Defect Lines and Fusion Category}

The fundamental objects responsible for the generalized symmetries in $d=2$ are the topological defect lines (TDLs).  
The conservation property of conventional symmetries is captured by the topological property of the TDLs that may join and split. 
Here we briefly review their key physical properties and set up the mathematical language of fusion category to describe them. We refer the readers to \cite{Fuchs:2002cm,Fuchs:2003id,
Frohlich:2004ef,Fuchs:2004dz,Fuchs:2004xi,Frohlich:2006ch,Frohlich:2009gb} for the original works on such symmetries in rational conformal field theories (RCFT) and related 3d TQFT, and \cite{Bhardwaj:2017xup,Chang:2018iay} for further details in general QFT. We will largely follow the conventions in the recent work \cite{Choi:2023xjw}.

\paragraph{Dual structure, direct sum and fusion product}
We denote the trivial TDL by $\id$. For each TDL $\cL(\Sigma)$ extended on a curve $\Sigma$, we define the dual TDL by its orientation reversal, namely
\ie 
 \overline\cL(\Sigma) =\cL(\overline\Sigma)\,,
\fe 
 as an operator equation where $\overline\Sigma$ has the opposite orientation compared to $\Sigma$. Note that a TDL can be self-dual (such as the identity TDL $\id$), i.e. $\overline\cL(\Sigma)=\cL(\Sigma)$.
The TDLs have a natural direct sum structure (denoted by $\oplus$) which amounts to 
\ie 
 \la (\cL_i\oplus \cL_j)(\Sigma) \cdots \ra  = \la (\cL_i(\Sigma) \cdots \ra +\la (\cL_j(\Sigma) \cdots \ra \,,
\fe
in general correlation functions. The indecomposable TDLs are also known as the simple TDLs. The extended operator product of a pair of simple TDLs inserted at $\Sigma$ and at $\Sigma_\ep$ which is slightly displaced transversely from $\Sigma$, defines the fusion product (denoted by $\otimes$), 
\ie 
(\cL_i \otimes \cL_j)(\Sigma)=\lim_{\ep \to 0} \cL_i(\Sigma_\ep) \cL_j(\Sigma)=\oplus_k N_{ij}^k \cL_k(\Sigma)\,, 
\label{fusionring}
\fe
 with fusion coefficients $N_{ij}^k \in \mZ_{\geq 0}$. The fusion product of the simple TDLs then generates a fusion ring. The simple TDLs are $\cL_i,{\overline \cL}_i$ such that $N_{i\bar i}^\id=1$\,.

 A finite set of simple TDLs $\{\cL_i\}$ that closes under the operation of dual, direct sum, and fusion defines a fusion category $\cC$ which is a symmetry of the underlying QFT. 
 The objects of $\cC$ are in one-to-one correspondence with these TDLs and the simple objects are the simple TDLs which we denote collectively as ${\rm Irr}(\cC)$.
 
 We emphasize that a fusion category $\cC$ in general describes a finite subcategory in the full symmetry category of the QFT $\cT$ which we denotes as $\cC_{\cT}$,
 \ie 
 \cC \subset \cC_{\cT}\,,
 \label{QFTsymcategory}
 \fe
where $\cC_{\cT}$ is generally a rigid $\mC$-linear monoidal category that is not finite.

\paragraph{Morphisms and topological junctions}
 Importantly, TDLs admit topological junctions, which correspond to morphisms in the fusion category. The topological junctions form a vector space over $\mC$ known as the junction vector space. It suffices to discuss the topological junctions among simple TDLs as the general case follows from the distributive nature of the direct sum. We assume there is no topological junction (end-point) for a nontrivial TDL. This is equivalent to requiring the corresponding symmetry to be \textit{faithful}, because otherwise the TDL could break, retract to nothing, and thus cannot be detected by physical observables. For two simple TDLs $\cL_i,\cL_j$, the junction vector space is ${\rm Hom}_\cC(\cL_i,\cL_j)\cong\mC$ if they are identical, otherwise it's empty.  
For three simple TDLs, the junction vector space between two ingoing TDLs and one outgoing TDL is denoted by ${\rm Hom}_\cC(\cL_i \otimes \cL_j,\cL_k)$. In particular the fusion coefficient satisfies $N^k_{ij}=\dim {\rm Hom}_\cC(\cL_i \otimes \cL_j ,\cL_k)$. 
 The dual structure implies the following obvious isomorphisms
 \ie 
 {\rm Hom}_\cC(\cL_i \otimes \cL_j,\cL_k) \cong 
  {\rm Hom}_\cC(\cL_j \otimes \overline\cL_k,\overline\cL_i) \cong 
  {\rm Hom}_\cC(\overline\cL_k \otimes \cL_i,\overline\cL_j) \cong 
  {\rm Hom}_\cC(\cL_i \otimes \cL_j\otimes \overline{\cL}_k,\id) \,,
 \fe 
which in particular implies 
\ie 
N_{ij}^k=N_{j\bar k}^{\bar i}=N_{\bar k i}^{\bar j}\,.
\fe
The operator product between junction vectors $v\in {\rm Hom}(\cL_i,\cL)$ and $u\in {\rm Hom}(\cL,\cL_j)$ is given by composition of the morphisms which we denote as $u \circ  v \in {\rm Hom}(\cL_i,\cL_j)$.\footnote{Here by ${\rm Hom}$ we mean ${\rm Hom}_\cC$. For notational simplicity, we will often drop the subscript on ${\rm Hom}$ when there is no confusion from the context.}

A general topological configuration for $\cC$ is constructed by gluing TDLs with topological junctions, which represents a generalized background gauge field configuration for the symmetry $\cC$. Observables in the $\cC$-symmetric QFT are invariant under isotopy (i.e. deformations of the TDL configuration that does not cross itself or other insertions). 

\paragraph{Associator and F-symbols} As mentioned in the introduction, a key ingredient of non-invertible symmetry is its intrinsically non-trivial F-symbols that encode a generalized version of the 't Hooft anomaly for conventional discrete group symmetries. In the context of the above discussion, F-symbols arise from the associativity structure of fusion, which specifies a particular isomorphism, known as the associator $\A_{i,j,k}$, between the fusion product of three simple TDLs $\cL_i,\cL_j,\cL_k$ in different orders,
\ie 
\A_{i,j,k} \in {\rm Hom}((\cL_i \otimes \cL_j)\otimes \cL_k,\cL_i \otimes (\cL_j\otimes \cL_k))\,.
\fe
More explicitly, after projecting the triple fusion product to simple TDLs $\cL_\ell$, this is captured by the F-symbol $F_{ijk}^{\ell}$ which can be represented as a matrix (known as the F-matrix) with components $[F_{ijk}^{\ell}]_{(p,\A,\B),(q,\C,\D)}$ determining the change of basis between the isomorphic four-point junction vector spaces ${\rm Hom}((\cL_i \otimes \cL_j)\otimes \cL_k,\cL_\ell)$ and ${\rm Hom}(\cL_i \otimes (\cL_j\otimes \cL_k),\cL_\ell)$, where $p,q$ label the simple TDLs $\cL_p\in \cL_i\otimes \cL_j,\cL_q \in \cL_j\otimes \cL_k$ in the fusion channels and $\A,\B,\C,\D$ label the independent junction vectors at the corresponding trivalent junctions. As usual, these matrix components of the F-symbol are basis dependent in general. Two sets of F-symbols $F_{ijk}^{\ell}$ are equivalent if and only if their matrix components are related by a change of basis at each trivalent junction vector space (i.e. there is a gauge freedom from $[M_{ij}^k]_{\A\B}\in {\rm GL}(N_{ij}^k,\mC)$ for each triplet of simple TDLs $\cL_i,\cL_j,\cL_k$).

Physically, the F-symbols relate different topological configuration of the TDLs via the so-called F-moves (i.e. 2-2 Pachner moves in the dual triangulation). In particular, one can perform local fusion of TDLs using the F-moves. The F-symbols obey nontrivial algebraic constraints known as pentagon equations, coming from the consistency condition under consecutive changes of basis in the five-point junction vector space (which can be derived from the isotopy invariance of the associated TDL configuration \cite{Chang:2018iay}). Unsurprisingly the pentagon equations are direct analogs of the cocycle conditions for group cohomology $H^3(G,U(1))$ that classifies anomalies for discrete group symmetry. Once the pentagon equations are solved, there are no further independent consistency conditions on the F-symbols, as a consequence of the Mac Lane coherence theorem \cite{etingof2016tensor}. The pentagon equations have a finite number of solutions, a result known as the Ocneanu rigidity and a given fusion ring may not even admit a single solution \cite{etingof2016tensor}. It is an open question to systematically classify fusion categories, which appears much richer than the classification of discrete groups.

\begin{figure}[!htb]
    \centering
    \includegraphics[scale=1.2]{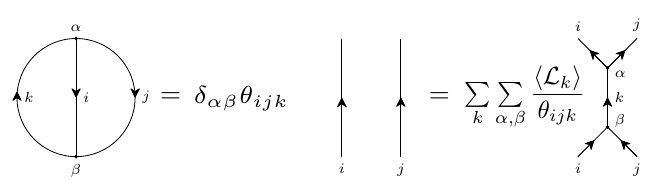}
    \caption{Positive configurations  of TDLs in the positive gauge ($\theta_{ijk}>0$) for the F-matrices.}
    \label{fig:Fpositivegauge}
\end{figure}

\paragraph{Adjoint structure and unitarity}
So far we have avoided dwelling upon an important property of the QFTs that we consider here, namely unitarity. Correspondingly, non-invertible TDLs in a unitary QFT are described by a unitary fusion category. At the level of the data for the fusion category introduced above, unitarity is encoded in the adjoint structure of the morphism spaces, which specifies an anti-linear map, 
\ie 
\dagger: {\rm Hom}_\cC(\cL,\cL')  \to {\rm Hom}_\cC(\cL',\cL)
\fe
for all $\cL,
\cL'\in \cC$. The map $\dagger$ defines a bilinear form via the correlation function $\la u^\dagger v\ra$ of junction vectors $u,v\in {\rm Hom}_\cC(\cL,\cL')$.
The fusion category is unitary if this bilinear form is positive definite, which follows from the reflection positivity of the underlying QFT. In particular, there exists a gauge (i.e. choice of basis in each trivalent junction vector space) such that the associator (equivalently F-symbols) is unitary
\ie 
\A_{ijk}\circ (\A_{ijk})^\dagger=\oplus_\ell \left( F_{ijk}^{\ell}\circ (F_{ijk}^{\ell} )^\dagger \right)  ={\rm id}_{(\cL_i\otimes \cL_j)\otimes\cL_k}\,,
\label{Funitary}
\fe 
where ${\rm id}_\cL$ the canonical identity morphism associated to $\cL\in \cC$ (physically this comes from the bulk identity operator restricted on the TDL $\cL$). In terms of the F-matrix, \eqref{Funitary} is equivalent to,
\ie 
[F_{ijk}^{\ell}]_{(p,\A,\B),(q,\C,\D)} ([F_{ijk}^{\ell'}]_{(q,\C,\D),(r,\ep,\zeta)})^*=\D_{\ell\ell'}\D_{pq}\D_{\A\ep}\D_{\B\zeta}\,.
\fe
The above does not completely fix the gauge freedom of the F-matrix in general. It is natural to impose the additional constraint that reflection symmetric TDL configurations are positive since they naturally compute the inner product in the corresponding junction vector space, which we refer to as the \textit{positive gauge} for the F-matrix (see Figure~\ref{fig:Fpositivegauge}).\footnote{By choosing a (conventional) gauge for the F-symbols (see e.g. \cite{Kitaev:2005hzj}), we have $\theta_{ijk}^2=\la\cL_i\ra\la\cL_j\ra\la\cL_k\ra$.}
We will work in this gauge unless explicitly stated otherwise.

\paragraph{Symmetry transformations, quantum dimensions, and defect Hilbert spaces}

The $\cC$-symmetric CFT enriches the algebraic structure of the symmetry by providing analytical data such as the operator spectrum and correlation functions that transform under the symmetry. In particular, shrinking a TDL $\cL$ encircling a local operator $\cO(x)$ produces another local operator $\hat \cL \cdot \cO(x)$ in the same spacetime representation. The symmetry transformation corresponding to the TDL $\cL$ is encoded in the linear map $\hat\cL$ which naturally acts on the CFT Hilbert space $\cH_{S^1}$ on $S^1$ by the operator-state correspondence, and it preserves the conformal multiplets labeled by the primary conformal weights $(h,\bar h)$ with respect to the left and right Virasoro symmetries.
The linear maps $\hat \cL_i$ for a collection of TDLs $\cL_i$ obviously obey the same fusion rules (on the cylinder). For a CFT with a unique conformally invariant vacuum, which is the focus here, the eigenvalue of $\cL_i$ with respect to the vacuum $|0\ra$, is the quantum dimension of $\cL_i$,
\ie 
\hat \cL_i |0\ra =\la \cL_i\ra |0\ra\,.
\fe
Equivalently the quantum dimension is the expectation value of an empty TDL loop on the plane (up to an isotopy anomaly \cite{Chang:2018iay,Cordova:2019wpi}).
It follows then that $\la \cL_i\ra$ obeys the fusion ring relation \eqref{fusionring}. In unitary theories, the quantum dimension is always bounded from below,
\ie 
\la \cL_i \ra \geq 1\,,
\fe
and the inequality is saturated if and only if $\cL_i$ is invertible \cite{Chang:2018iay}.

By the locality of the underlying CFT, a TDL $\cL$ extended in the (Euclidean) time direction defines a new Hilbert space $\cH_{S^1}^\cL$, that generalizes the conventional twisted sector for discrete group symmetries. By the state-operator correspondence in the presence of the TDL, states in $\cH_{S^1}^\cL$ are mapped to $\cL$-twisted sector operators, namely operators on which $\cL$ can end. For nontrivial $\cL$, by the faithfulness condition (vanishing tadpole condition in \cite{Chang:2018iay}), the ground states in $\cH_{S^1}^\cL$ necessarily have conformal weight $h+\bar h>0$, and the spin spectrum contains nontrivial information about the F-symbols involving $\cL$ \cite{Chang:2018iay}. The latter can be derived by studying how symmetries act in the $\cL$-twisted sector via configurations involving other TDLs along the spatial direction. The defect Hilbert space $\cH_{S^1}^\cL$ generalizes in the obvious way for multiple parallel TDLs in the temporal direction which we denote as $\cH_{S^1}^{\cL_1\cL_2\dots \cL_n}$. Now it is possible to have a topological operator, namely $h=\bar h=0$, in the twisted sector, and the $h=\bar h=0$ subspace of $\cH_{S^1}^{\cL_1\cL_2\dots \cL_n}$ is the topological junction vector space ${\rm Hom}_\cC(\cL_1\otimes \cL_2\otimes \cdots \otimes \cL_n,\id)$.

\subsection{Algebra Objects and Generalized Gauging}

We now review several key mathematical notions related to fusion category together with their physical incarnations. We will use them in the ensuing sections to describe the generalized gauging of fusion category symmetries in QFT.

\begin{figure}[!htb]
    \centering
   \includegraphics[scale=1]{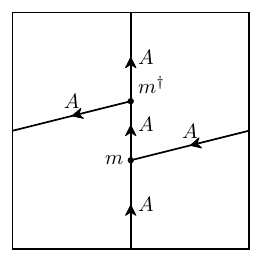}
    \caption{Discrete gauging on the torus.}
    \label{fig:torus}
\end{figure}

 Recall that gauging a non-anomalous discrete group symmetry $G$ amounts to coupling the $G$-symmetric QFT $\cT$ to $G$ background gauge fields and performing a weighted sum over distinct configurations of the gauge fields. In the language of TDLs, this corresponds to inserting a topological network of invertible TDLs $\cL_g$ labeled by $g\in G$, which is equivalent to a flat $G$ background as $\cL_g$ specifies the transition function from one patch to another in the dual triangulation. Yet another equivalent way to view this is to consider the non-simple TDL $A=\oplus_{g\in G} \cL_g$ and insert a fine-enough mesh of $A$ made of trivalent topological junctions of a specific type $m$ (and $m^\dagger$) among $A$ on the spacetime manifold. In particular, the usual $G$-orbifold at the level of the torus partition function can be represented as in Figure~\ref{fig:torus},
where the topological junction $m$ is not unique in general and physically inequivalent choices of $m$ are labeled by the discrete torsion $H^2(G,U(1))$ (i.e. the weighting mentioned above in the sum over gauge field configurations).

\begin{figure}[!htb]
    \centering
     \includegraphics[scale=0.54]{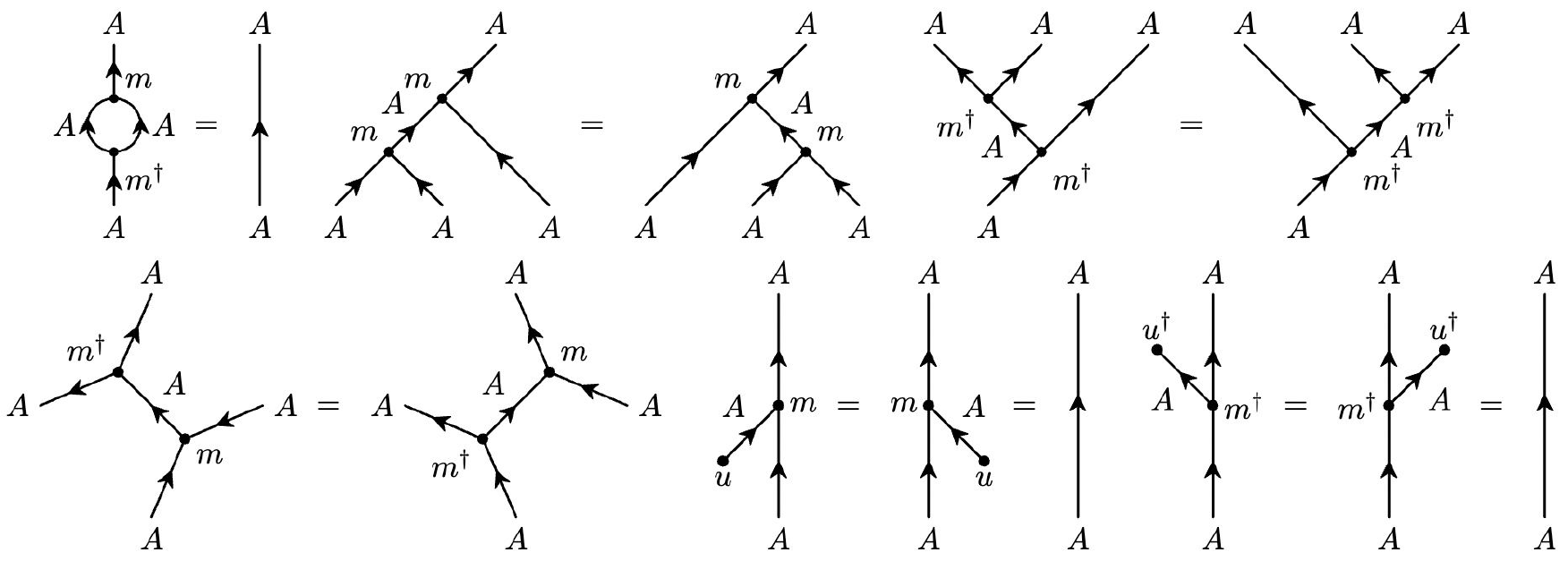}
    \caption{The separability, associativity, Frobenius, and unit conditions on the algebra object $(A,m,m^\dagger,u,u^\dagger)$. Here we have written them in the positive gauge. In the general gauge, the comultiplication morphism $m^\dagger$ and the counit $u^\dagger$ should be replaced by $m^\vee$ and $u^\vee$ to avoid confusions.}
    \label{fig:algebraconditions}
\end{figure}

It is this last description that generalizes immediately to gauging TDLs in a  general fusion category symmetry $\cC$. The basis data that specifies the background gauge field consists of a TDL $A\in \cC$ which is non-simple in general and a trivalent junction $m\in {\rm Hom}_\cC(A\otimes A,A)$ known as the multiplication morphism which also gives rise to the co-multiplication morphism $m^\dagger\in {\rm Hom}_\cC(A,A\otimes A)$ by the adjoint structure.   
What replaces the anomaly-free condition in the case of a discrete group $G$ are the diagrammatic equalities known as the separability and associativity conditions in Figure~\ref{fig:algebraconditions},
which are direct analogs of the gauge invariance conditions for usual gauging and generate all topological moves connecting two different TDL configurations \cite{Turaev:1992hq}.\footnote{In $d=2$, the two topological moves in Figure~\ref{fig:algebraconditions} are known as the Matveev moves \cite{Turaev:1992hq}.} Furthermore, to ensure the vacuum survives after gauging, we require the algebra object $A$ to have a unit $u\in {\rm Hom}_\cC(\id, A)$ that satisfies the unit condition (similarly for the co-unit $u^\dagger\in {\rm Hom}_\cC(A,\id)$) in Figure~\ref{fig:algebraconditions}.

The full set of algebraic conditions on $(A,m,m^\dagger,u,u^\dagger)$ that ensures the consistency of gauging defines a symmetric separable special Frobenius algebra object in $\cC$ \cite{Fuchs:2002cm}.\footnote{Unitarity is not required to define a symmetric separable special Frobenius algebra object in a general fusion category. In this more general scenario, the defining data for the algebra is $(A,m,m^\vee,u,u^\vee)$ where $m^\vee$ denotes the co-multiplication morphism and $u^\vee$ is the co-unit and they satisfy the same set of algebraic constraints including those in Figure~\ref{fig:algebraconditions}. We emphasize that even for algebras in a unitary fusion category, if the F-matrices are not in the positive gauge the co-multiplication morphism $m^\vee$ can differ from $m^\dagger$.} We will simply refer to them as algebra objects in this work and denote them as $(A,m)$ or simply as $A$ when there is no room for confusion. Note that in any fusion category, there is a trivial algebra object corresponding to $A=\id$, which physically means trivial gauging.

\subsection{Half-gauging, Topological Gauging Interfaces and Self-duality}
\label{sec:halfgauging}

\begin{figure}[!htb]
    \centering
   \includegraphics[scale=2]{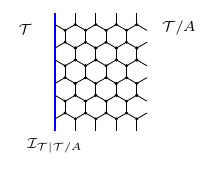}
    \caption{Topological gauging interface from half-gauging algebra object $A\in \cC$.}
    \label{fig:halfgauging}
\end{figure}

As in the case of discrete group symmetries, instead of gauging an algebra $A\in \cC$ for a symmetry $\cC$ of a QFT $\cT$ on the entire spacetime, we can consider gauging on a submanifold $\cS$ of the spacetime. In particular, by choosing a topological boundary condition for the relevant gauge fields at the boundary $\pa \cS$, this procedure produces a topological interface $\cI_{\cT|\cT/A}(\cS)$ (which can be disconnected if $\cS$ is) located at $\cS$ interpolating between the original theory $\cT$ and its gauged cousin $\cT/A$. 
Due to the topological nature of this interface, to study its properties it suffices to focus on the simplest setup where $\cS$ is half of the spacetime as in Figure~\ref{fig:halfgauging}.

\begin{figure}[!htb]
    \centering
   \includegraphics[scale=2]{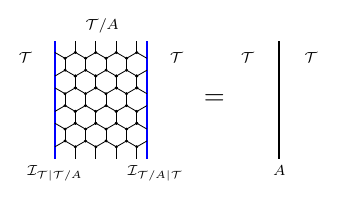}
    \caption{Fusion of topological gauging interfaces from half-gauging $A$.}
    \label{fig:interfacefusion}
\end{figure}

Here, as explained in the last subsection, the sum over gauge field configurations for $A\in \cC$ is represented by a fine-enough mesh of $A$ on $\cS$ built from their topological junctions $m$ (and its adjoint $m^\dagger$). The topological boundary condition at $\pa \cS$ is simply given by the insertion of $A$ itself. One can consider the fusion of the topological interface $\cI_{\cT|\cT/A}$ and its orientation reversal $\cI_{\cT/A|\cT}$ as in Figure~\ref{fig:interfacefusion}, which produces gauging on the internal strip and consequently the following simple fusion product,
\ie 
\cI_{\cT|\cT/A} \otimes \cI_{\cT/A|\cT}=A\,.
\label{interfacefusion}
\fe
Note that the fusion of the topological interface $\cI_{\cT|\cT/A}$ with TDLs in $\cC$ from the left (and similarly for $\cI_{\cT/A|\cT}$ from the right) produce other topological interfaces between $\cT$ and $\cT/A$ (see Section~\ref{sec:modulecat} for more discussions).

Furthermore if the region $\cS$ is compact and contractible (e.g. a disk), one can consider shrinking the topological interface enclosing $\cS$ down to nothing, which produces a constant multiplying the identity operator (since $\cT$ and $\cT/A$ have a unique vacuum), and this constant defines the quantum dimension of the interface.\footnote{It coincides with the $g$-function for general conformal interfaces among 2d CFTs and is positive by unitarity (in fact bounded from below by 1).} Note that the quantum dimension of the interfaces are invariant under orientation reversal. Together with the fusion rule \eqref{interfacefusion}, it follows that 
\ie 
d_{\cI_{\cT|\cT/A}}=d_{\cI_{\cT/A|\cT}}=\sqrt{\la A \ra}\,.
\label{interfaceqdim}
\fe
This can be derived from \eqref{interfacefusion} near the equator of $S^2$ by shrinking the two interfaces towards the two opposite poles, respectively, and using the fact that the theories $\cT$ and $\cT/A$ have identical partition functions on $S^2$, which has no nontrivial one-cycles.\footnote{For topological interfaces that are self-dual (this can happen for $\cI_{\cT|\cT/A}$ if $\cT\cong \cT/A$), there is a notion of interface Frobenius-Schur indicator which can be detected from the ``boundary crossing relation'' in \cite{Huang:2021zvu} and in a related way by the module trace 
\cite{Schaumann_2013}.} 

As is the case for any interfaces in QFT, an equivalent way to think about the topological gauging interface is to use the folding trick and consider boundary conditions of the folded QFT, here given by special boundary conditions of $\cT\times \overline \cT/A$. In $d=2$ CFTs, there is a canonical way to construct such a boundary condition using a generalization of the regular brane for a discrete symmetry group $G$ \cite{Thorngren:2021yso}, and the fusion rule \eqref{interfacefusion} then follows from consideration of the cylinder partition function with this boundary condition at the two ends.

One can also consider the fusion of the topological gauging interface and its orientation reversal in the opposite way,
\ie 
\cI_{\cT/A|\cT} \otimes \cI_{\cT|\cT/A}=A^*\,,
\label{interfacefusiondual}
\fe
producing a (non-simple) TDL in the theory $\cT/A$, which we denote as $A^*$, that must have the same quantum dimension as $A$
\ie 
\la A^*\ra=\la A \ra\,.
\label{AdualAqdim}
\fe 
As we explain in the next subsection, $A^*$ naturally defines the dual algebra object such that gauging $A^*$ brings back the original theory,
\ie 
\cT/A/A^*\cong \cT\,.
\fe

An interesting case is when the gauged theory is isomorphic to the original theory,
\ie 
\cT \cong \cT/A\,,
\label{selfdualgauging}
\fe
then the topological gauging interface becomes a topological defect in the same QFT $\cT$. In particular, it corresponds to a TDL $\cN$ and its dual $\overline{\cN}$,
\ie 
\cN=\cI_{\cT|\cT/A}\,,\quad \overline{\cN}=\cI_{\cT/A|\cT}\,,
\fe
with fusion rules
\ie 
&\cN\otimes\overline{\cN}=A\,,\quad \overline{\cN}\otimes {\cN}=A^*\,,\quad 
\label{dualityTDLfusion}
\fe
where the isomorphism in \eqref{selfdualgauging} is implicitly used to identify TDLs in $\cT/A$ with those in $\cT$.\footnote{The full set of fusion rules involving the simple TDLs in $A$ and $A^*$ with $\cN$ and $\overline{\cN}$ will be captured by the category of right and left $A$-modules, respectively (physically these correspond to other possible simple topological interfaces between $\cT$ and $\cT/A$). See Section~\ref{sec:modulecat} for related discussions.}
Note that, in general $A$ and $A^*$ are non-isomorphic as TDLs.
Furthermore, the F-symbols involving the TDL $\cN$ (and its dual) will depend on the multiplication morphism $m$ for the corresponding algebra $A$ that implements the self-dual gauging.

Conversely, if a QFT $\cT$ admits TDLs $\cN,\overline{\cN}$ that obey the fusion rules in \eqref{dualityTDLfusion}, $A$ is guaranteed to be an algebra object with a unique multiplication morphism determined by the F-symbols involving $\cN,\overline{\cN}$ (similarly, there is an algebra object associated to $A^*$). The QFT then satisfies self-duality under gauging either $A$ or $A^*$. Intuitively, this follows from the fact that an arbitrary $A$-mesh can be constructed by starting from a contractible oriented loop of $\cN$, deforming it and bring it around various nontrivial cycles of the spacetime manifold. Therefore, the $A$-mesh produces identical observables in $\cT$ as is the case without it (similarly for the $A^*$-mesh). 

The above generalizes the self-dual gauging for a discrete abelian group symmetry $G$, in which case the TDL $\cN$ is self-dual $\cN\cong \overline{\cN}$ (consequently $A\cong A^*$ in \eqref{dualityTDLfusion}) and generates a well-studied $\mZ_2$ extension of the group category ${\rm \bf Vec}_G$ known as the Tambara-Yamagami fusion category ${\rm \bf TY}(G,\chi,\ep)$, where the F-symbol, up to equivalence relations, is entirely determined by a symmetric non-degenerate bicharacter $\chi:G\times G\to U(1)$ and the Frobenius-Schur indicator $\ep=\pm 1$ for the self-dual TDL $\cN$ \cite{tambara1998tensor,tambara2000representations}. Here $\cN$ explains the self-duality of the QFT with TY symmetry under $G$-gauging, and thus $\cN$ is often referred to as the duality TDL (for $G$-gauging). Here we see a vast generalization of such duality TDLs which arise when considering non-invertible gauging.

To summarize, we have shown that 
\begin{theorem}
 A $\cC$-symmetric QFT $\cT$ is self-dual under gauging an algebra object $A\in \cC$ if and only if $\cT$ admits a duality TDL $\cN$ (and its dual $\overline{\cN}$) with fusion rules \eqref{dualityTDLfusion}. 
 \label{thm:selfdual}
\end{theorem}

Note that either $\cN\in \cC$ or the QFT $\cT$ must have a large symmetry (sub)category $\cC'$ that extends $\cC$ by $\cN$ (in any case $\cN\in \cC_{\cT}$ defined in \eqref{QFTsymcategory}). 
In the former case, one easily comes up with numerous examples of self-dual gauging from any self-dual TDL in any fusion category $\cC$ that describes a subcategory of the full symmetry category of the QFT $\cT$ (see Section~\ref{sec:Isingsq}). The simplest nontrivial example is perhaps $A=\id \oplus W$ in the Fib category with two simple TDLs $\{\id,W\}$; this category describes the symmetries of many CFTs, including the tri-critical Ising CFT. The corresponding self-duality defect is simply $\cN=W$. Alternatively, this provides a way to discover TDLs in the potentially vast symmetry category $\cC_\cT$ of a given $\cT$ that exhibits self-duality under gauging. For example, this is how various self-duality defects for invertible gauging were discovered in irrational CFTs in \cite{Thorngren:2019iar,Thorngren:2021yso,Nagoya:2023zky} where one lacked the algebraic tools to directly identify the full symmetry category.

As we explain in Section~\ref{sec:modulecat}, Theorem~\ref{thm:selfdual} also means that any self-dual gauging in QFT $\cT$ produces a Morita trivialization of the corresponding algebra object $A$ in a subcategory of the full symmetry category of $\cT$.

\subsection{General Topological Interfaces and (Bi)module Categories}
\label{sec:modulecat}

In the last subsection, we saw that performing half-gauging using an algebra object $A$ in the $\cC$-symmetric theory $\cT$ with Dirichlet boundary conditions produced a distinguished topological gauging interface $\cI_{\cT|\cT/A}$ between the original theory $\cT$ and its gauged cousin $\cT/A$. In fact, as we explain below, any topological interface between two QFTs $\cT$ and $\cT'$ corresponds to a topological gauging interface for an algebra object $A$ in the symmetry category of $\cT$ (and similarly for an algebra object $A^*$ in the symmetry category of $\cT'$). In other words, we establish the following simple theorem\footnote{The connection between discrete gaugings and topological interfaces was already pointed out in \cite{Bhardwaj:2017xup} which translates related results in category theory into the QFT language. Here we emphasize the physical picture which involves the fusion of topological interfaces among themselves and also with bulk TDLs.}
\begin{theorem}
Two QFTs $\cT$ and $\cT'$ are related by discrete generalized gauging if and only if there exists a topological interface $\cI_{\cT|\cT'}$ between $\cT$ and $\cT'$. The corresponding algebra object in the symmetry category $\cC$ of $\cT$ comes from the fusion of the topological interface and its dual, namely $A=\cI_{\cT|\cT'}\cI_{\cT'|\cT}$.
 \label{thm:gauginginterface}
\end{theorem}

\begin{figure}[!htb]
    \centering
    \includegraphics[scale=1.3]{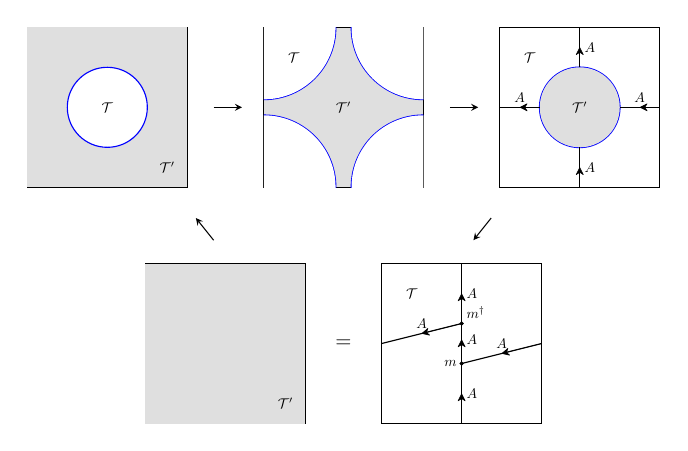}
    \caption{A sequence of topological moves that relates a topological interface to gauging on the torus.}
    \label{fig:interfacetogauging}
\end{figure}

The fact that $A=\cI_{\cT|\cT'}\cI_{\cT'|\cT}$ defines an algebra object simply follows from the topological property of the interface $\cI_{\cT|\cT'}$. Starting from a contractible bubble of $\cI_{\cT|\cT'}$ separating $\cT$ from  $\cT'$, one can freely deform its shape by stretching and pinching its sides to produce solutions to the defining equations for an algebra object. The above theorem follows immediately. In Figure~\ref{fig:interfacetogauging}, we illustrate how this works on the torus by deriving a relation between the partition functions of the two theories $\cT,\cT'$, as expected from generalized gauging. Below we elaborate on this perspective and
describe general features of generalized gauging in terms of the topological interfaces.

\paragraph{Module categories as categories of topological interfaces}
As mentioned in Section~\ref{sec:halfgauging}, topological interfaces naturally come in families, obtained from fusing TDLs from both left and right. Furthermore, there is a natural direct sum structure on the set of topological interfaces between two QFTs $\cT,\cT'$ which respects the locality property of these extended objects. In particular, each pair of topological interfaces $\cI_1$ and $\cI_2$ between $\cT$ and $\cT'$ is associated with a Hilbert space $\cH^{\cI_1\overline{\cI}_2}_{S^1}$ on a transverse $S^1$ bi-partitioned into two segments, and the direct sum structure on the interfaces naturally corresponds to the direct sum structure on this Hilbert space. By the folding trick, each interface is equivalent to a boundary condition for the product theory $\cT\times \overline{\cT'}$, and states in the  above Hilbert space naturally map to states in the Hilbert space of the product theory on a strip with the corresponding boundary condition at the two ends. In CFT, the states in $\cH^{\cI_1\overline{\cI}_2}_{S^1}$ correspond to interface-changing operators between $\cI_1$ and $\cI_2$. As for the TDL twisted Hilbert space, the topological operators are captured by the $h=\bar h=0$ subspace of $\cH^{\cI_1\overline{\cI}_2}_{S^1}$, and the simple (irreducible) interfaces are those that admit a unique topological operator (up to rescaling) on their worldvolumes from the identity operator in the bulk. 

Let us define $\{\cI_a\}$ as a set of simple topological interfaces between $\cT$ and $\cT'$ that close under fusion with simple TDLs $\cL_i$ representing symmetry $\cC$ of $\cT$. Consistency with locality and fusion requires $\cL_i$ to act linearly via non-negative integer matrix representations (NIM-reps \cite{Bockenhauer:1999wt}) of the fusion ring for $\cC$,\footnote{See \cite{Behrend:1999bn,Gannon:2001ki} where NIM-reps in RCFTs (for Cardy branes) have been studied extensively.}
\ie 
\cL_i \otimes \cI_a = \oplus_b P_{ia}{}^b \cI_b\,,\quad P_i \cdot P_j = \sum_k N_{ij}^k P_k\,,
\label{NIMrepdef}
\fe
where $P_{ia}{}^b$ has non-negative integer entries and counts the number of independent topological junctions between $\cL_i,\cI_a$ and $\cI_b$. There is an analog of the F-matrix for the change of basis matrix in junction vector spaces between bulk TDLs and interfaces.
Consistency from isotopy invariance in the presence of multiple topological junctions along an interface leads to an interface analog of the pentagon equation (see \cite{Choi:2023xjw} for a recent review). The full mathematical structure (up to equivalence from basis change at the junction vector spaces) is captured by a left $\cC$-module category, which we denote as $\cM\in {\rm Mod}(\cC)$.  In this language, the simple interfaces $\cI_a$ are the simple objects in $\cM$, the topological junction vector space between two general interfaces $\cI,\cI'$ (from direct sums of $\cI_a$) is identified with the Hom space ${\rm Hom}_\cM(\cI,\cI')$. In particular, the NIM-rep gives 
\ie 
P_{ia}{}^b=\dim {\rm Hom}_\cM(\cL_i\otimes \cI_a,\cI_b)\,.
\fe
Without loss of generality, we focus on indecomposable (simple) module categories $\cM$ in which case the corresponding NIM-rep $\{P_i\}$ is also indecomposable. Physically, the $\cC$-module category $\cM$ is the category of topological interfaces between $\cT$ and $\cT'$ that close under fusion with TDLs in $\cC$.

\begin{figure}[!htb]
    \centering
   \includegraphics[scale=1.55]{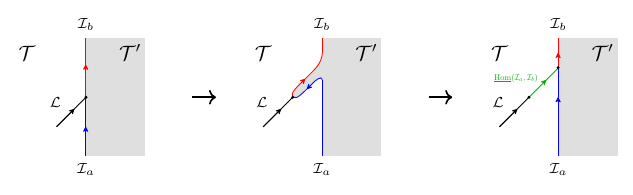}
    \caption{Internal Hom space from interface fusion.}
    \label{fig:internalhom}
\end{figure}

\paragraph{Algebra objects from interface fusion and internal Hom}
For each simple interface $\cI_a$ between the theories $\cT$ and $\cT'$, the corresponding algebra object $A\in \cC$ that relates the two by discrete gauging follows from the interface fusion, 
 \ie 
A=\cI_a \otimes \overline{\cI}_a\,,\quad \cT'\cong \cT/A\,,
\label{Afrominterfacefusion}
 \fe
according to Theorem~\ref{thm:gauginginterface}. This gives a physical picture of the internal Hom $\underline{\rm Hom}(\cdot,\cdot)\in \cC$ introduced in \cite{ostrik2001module} and defined by the isomorphism,  
\ie
{\rm Hom}_\cM(\cL\otimes \cI_a,\cI_b)\cong {\rm Hom}_\cC(\cL,\underline{\rm Hom}(\cI_a,\cI_b))\,,
\fe
for general TDL $\cL\in \cC$. In general, the internal Hom is a non-simple TDL obtained from the interface fusion (see Figure~\ref{fig:internalhom}),
\ie 
\underline{\rm Hom}(\cI_a,\cI_b) = \cI_b \otimes \overline{\cI}_a\,,
\label{inthomfusion}
\fe
and \eqref{Afrominterfacefusion} is a special case.

The algebra objects obtained from fusing a simple topological interface and its dual in \eqref{Afrominterfacefusion} contain the trivial TDL without degeneracy,
\ie 
{\rm Hom}_\cC(\id, A)\cong \mC\,,
\fe
and are known as haploid (or connected) algebras.\footnote{Note that haploid is a stronger condition than simplicity on an algebra object $A$. The latter only requires $A$ to be simple as a bimodule over itself whereas the former requires $A$ to be simple as a left-module over itself. One simple example of an algebra object that is simple but not haploid is a finite dimensional matrix algebra ${\rm Mat}_n(\mC)$ in the category ${\rm \bf Vec}$ of finite dimensional vector spaces over $\mC$.} In the rest of the paper, all algebras are haploid unless otherwise noted explicitly.\footnote{Physically gauging haploid algebras preserve the condition of having a unique vacuum in the QFT.} The algebras from different simple interfaces $\cI_a\in \cM$ (also for non-simple interfaces) in the $\cC$-module category give rise to physically equivalent gaugings of TDLs in $\cC$.

Conversely, given an algebra object $A\in \cC$ in the symmetry category of $\cT$, one can reconstruct the topological interfaces between $\cT$ and $\cT/A$ and thus the correspondingly $\cC$-module category, in a canonical way. As aforementioned, general observables in the gauged theory $\cT/A$ are constructed from observables in the original theory $\cT$ decorated by a fine-enough $A$-mesh. Topological interfaces between $\cT$ and $\cT/A$ are then constructed by configurations of the $A$-mesh ending topologically on TDLs $\cL_i\in \cC$ in $\cT$ from the right\footnote{Here we focus on the subset of topological interfaces that arise universally from the symmetry category $\cC$ alone. The theory $\cT$ may (and typically does) admit a larger symmetry than $\cC$ and correspondingly there are more topological interfaces to other theories. } in a way that respects the algebra structure of $A$, which ensures the invariance under topology changes in the $A$-mesh. Such topological interfaces have the natural structure of right $A$-modules in $\cC$, and they may admit topological interface changing operators that respect the $A$-module structure. 
Together they define the category $\cC_A$ of right $A$-modules in $\cC$, which admits a natural action from TDLs $\cL_i\in \cC$ by fusion from the left, and is equivalent to the left $\cC$-module category $\cM$ \cite{ostrik2001module}. The simple objects in the module category $\cM$ correspond to simple $A$-modules.
In particular, the algebra object $A$ itself is the obvious $A$-module which corresponds to the topological gauging interface discussed in Section~\ref{sec:halfgauging} by half-gauging $A$ with the Dirichlet boundary condition and is identified with the simple object $\cI_a \in \cM$ by \eqref{Afrominterfacefusion}. General simple $A$-modules are identified with general simple objects $\cI_b \in \cM$ via the internal Hom $\underline{\rm Hom}(\cI_a,\cI_b)$.
Any two algebra objects $A,B\in \cC$ that produce an isomorphism $\cC_A \cong \cC_B $ as $\cC$-module categories are said to be Morita equivalent in $\cC$ and lead to physically equivalent gaugings of TDLs in $\cC$ (in which case $A$ and $B$ can be recovered by the internal Hom from objects in $\cC_A$). The Morita equivalence classes of algebra objects in $\cC$, namely the physically distinct gaugings, are precisely captured by the inequivalent $\cC$-module categories. In Section~\ref{sec:halfgauging} (see Theorem~\ref{thm:selfdual}), we have seen that an algebra $A$ is Morita trivial (equivalent to $A=\id$) in $\cC$ if and only if $A=\cN\otimes \overline{\cN}$ for a TDL $\cN\in \cC$ which comes with the canonical algebra structure. More generally, a sufficient condition for two algebras $A,B\in \cC$ to be Morita equivalent is the existence of a TDL $\cN\in \cC$ such that
\ie 
B= \cN\otimes A\otimes \overline{\cN}\,,
\label{moritaequivalent}
\fe
again with a canonical algebra structure induced from that of $A$ on the RHS \cite{Roumpedakis:2022aik}.

\paragraph{Symmetries after gauging and the dual fusion category from the category of $(A,A)$-bimodules}
A subset of the TDLs in the gauged theory $\cT/A$ can be determined simply from the gauging procedure algebraically, and they generate a finite subcategory ${}_A\cC_A\subset \cC_{\cT/A}$ known as the dual category of $\cC$ with respect to $A$. They come from TDLs $\cL\in \cC$ with $A$-mesh ending topologically from either sides. Consistency conditions from invariance under topology change in the $A$-mesh with topological junctions between $\cL$ and $A$ (from either side) produce  $(A,A)$-bimodules. Such $(A,A)$-bimodules have a natural tensor product structure $\otimes_A$ and a corresponding associator, together defining a fusion category ${}_A \cC_A$ via the category of $(A,A)$-bimodules, which captures the dual symmetry in the gauged theory $\cT/A$ and generalizes the quantum symmetry $\hat G$ in an ordinary orbifold by a finite abelian group $G$. Intuitively, we refer to ${}_A \cC_A$ as the dual category to $\cC$ with respect to the algebra object $A\in \cC$. As expected from the above discussion, the dual category ${}_A \cC_A$ only depends on $A$ via its Morita equivalence class which is specified by the module category $\cC_A$. Thus we can also say that ${}_A \cC_A$ is dual to $\cC$ with respect to $\cC_A$. Furthermore $\cC_A$, which describes the category of topological interfaces between $\cT$ and $\cT/A$, has a natural action by fusing TDLs in ${}_A \cC_A$ from the right and thus the structure of a $(\cC,{}_A \cC_A)$-bimodule category.

\paragraph{Bimodule categories and categorical Morita equivalence}
More generally, a $(\cC,\cC')$-bimodule category for fusion categories $\cC$ and $\cC'$ represents a category of topological interfaces where TDLs in $\cC$ and $\cC'$ can end topologically from the left and from the right respectively. The bimodule categories have a natural tensor product from the fusion of the topological interfaces, which is mathematically defined as $\cM \boxtimes_{\cC'}   \cN$ for a $(\cC,\cC')$-bimodule category $\cM$ and a $(\cC',\cC'')$-bimodule category $\cN$ \cite{etingof2009fusion}. A $(\cC,\cC')$-bimodule category $\cM$ is called invertible if $\cM^{\rm op} \boxtimes_{\cC}   \cM \cong \cC'$ (equivalently $\cM \boxtimes_{\cC'}  \cM^{\rm op} \cong \cC$) as bimodule categories where $\cM^{\rm op}$ is the $(\cC',\cC)$-bimodule category defined by orientation reversal. Physically, any simple object in an invertible bimodule is a topological interface $\cI$ such that the fusion product contains the trivial TDL without degeneracy $\cI\otimes \overline{\cI}\ni \id_\cC$ and $\overline{\cI} \otimes \cI \ni \id_{\cC'}$. When this happens, the two fusion categories $\cC$ and $\cC'$ are said to be categorically Morita equivalent \cite{etingof2009fusion}.
The $(\cC,{}_A \cC_A)$-bimodule category $\cM=\cC_A$ obtained from gauging is precisely of this type, where $\cI$ is the topological gauging interface for the haploid algebra $A=\cI \otimes\overline{\cI}$ and $A^*=\overline{\cI}\otimes \cI$ is the dual algebra object. The $A$-gauging can be undone by further gauging $A^*$. In fact, two fusion categories $\cC$ and $\cC'$ are categorically Morita equivalent if and only if each is the dual of the other by gauging.

The categorical Morita equivalence is in fact a 2-equivalence between certain 2-categories \cite{MUGER200381,etingof2009fusion,grossmansnyderAH,etingof2016tensor}. Here the 2-category ${\rm Mod}(\cC)$ contains as objects module categories $\cM,\cN$ over a fusion category $\cC$, the 1-morphisms are given by $\cC$-module functors ${\rm Fun}_\cC(\cM,\cN)$, and the 2-morphisms by the natural transformations of $\cC$-module functors. Equivalently, the simple objects in  ${\rm Mod}(\cC)$ can be thought of as the set of haploid algebras $A\in \cC$ (referred to as division algebras in \cite{grossmansnyderAH}) up to Morita equivalence, while the 1-morphisms and the 2-morphisms are given by objects and morphisms respectively in the category of $(A,B)$-bimodules. 
In physical terms, the objects $\cM,\cN$ label distinct gaugings of TDLs in the symmetry subcategory $\cC$ of the theory $\cT$, the 1-morphisms correspond to topological interfaces between (potentially) different gauged theories, and the 2-morphisms describe topological interface changing operators. This 2-equivalence implies a bijection $\cN \to {\rm Fun}_\cC(\cM,\cN)$ between the module categories over a pair of fusion categories $\cC$ and ${}_A \cC_A$ that are dual with respect to the $\cC$-module category $\cM$. In particular, for $\cN=\cM$, this implies the isomorphism ${}_A \cC_A\cong (\cC^*_\cM)^{\rm op}$ as fusion categories with $\cC^*_\cM\equiv {\rm Fun}_\cC(\cM,\cM)$.

\subsection{Sequential Gauging, Generalized Orbifold Groupoid, and Module Categories for Group-theoretical Fusion Categories}
\label{sec:seqgauginggrouptheoretical}

The physical picture of arbitrary discrete generalized gauging represented by a topological interface between a pair of QFTs makes it clear that sequential gauging simply follows from the fusion of the corresponding topological interfaces. Up to Morita equivalence (which keeps track of physically distinct gaugings), this is captured by the Deligne tensor product of the corresponding invertible bimodule categories. For example, let's consider the sequential gauging by an algebra object $A\in \cC$ and then by another algebra object $B\in \cC'$ in the dual category $\cC'={}_A \cC_A$ which produces another dual category $\cC''={}_B \cC'_B$. Let's denote the corresponding invertible bimodule categories in the two steps by $\cM$ and $\cN$, respectively.
Then this sequential gauging 
is equivalent to the one-step gauging using the invertible $(\cC,\cC'')$-bimodule category $\cM \boxtimes_{\cC'} \cN$ and the corresponding algebra object in $\cC$ follows from interface fusion as explained in Section~\ref{sec:modulecat}.

\paragraph{Generalized orbifold groupoid and Brauer-Picard groupoid $\underline{\underline{\rm \bf BrPic}}$}
The fusion categories related by discrete gauging in this way form a graph where each connected component consists of nodes labeled by fusion categories that are categorically Morita equivalent and  connecting edges corresponding to inequivalent invertible bimodule categories. This is the generalized version of the orbifold groupoid introduced in \cite{Gaiotto:2020iye} where it was studied in detail for invertible gaugings.\footnote{A slight difference with \cite{Gaiotto:2020iye} is that self-dual gaugings do not create a separate node in the definition here (and in \cite{etingof2016tensor}).}
This generalized orbifold groupoid structure is mathematically captured by the Brauer-Picard groupoid $\underline{\underline{\rm \bf BrPic}}$; a 3-groupoid which is a special case of a 3-category whose objects are fusion categories, 1-morphisms are invertible $(\cC,\cC')$-bimodules between fusion categories $\cC$ and $\cC'$, 2-morphisms are equivalences between such invertible bimodules, and 3-morphisms are isomorphisms of such equivalences (see \cite{etingof2016tensor} for more details). Here we focus on the truncated 1-groupoid which is a connected subgroupoid that contains a seed fusion category $\cC$
which is denoted simply as ${\rm \bf BrPic}$ (where the $\cC$ dependence is implicit \cite{etingof2016tensor}). Obviously, the definition of this groupoid is independent of the fusion category up to categorical Morita equivalence. Physically, the groupoid ${\rm \bf BrPic}$ keeps track of the fusion structure of topological interfaces described by invertible bimodule categories over fusion categories that are related by discrete gauging (generalized orbifold) to $\cC$. As we will illustrate in examples in Section~\ref{sec:CFTrealization}, this immediately predicts the fate of sequential gauging in a $\cC$-symmetry theory $\cT$ without knowing details of the dynamical content of the theory (i.e. it tells when seemingly different gauging sequences are equivalent). In particular, the Morita auto-equivalences of $\cC$ (i.e. invertible $(\cC,\cC)$-bimodule categories) form a group denoted by ${\rm \bf BrPic}(\cC)$ and captures potential nontrivial isomorphisms of $\cC$.\footnote{We emphasize that while the underlying QFT $\cT$ retains the $\cC$-symmetry after the discrete gauging which corresponds to the $(\cC,\cC)$-bimodule category, in general $\cT$ is not necessarily self-dual under this gauging.} Such nontrivial isomorphism can come from an automorphism of the fusion category (e.g. a permutation of TDLs that respect the fusion rules) or stacking with a topological counter-term, equivalently an SPT for $\cC$.

 \paragraph{Module categories for group-theoretical fusion categories}
The groupoid structure from categorical Morita equivalence can also be used to understand the module categories of fusion categories from a small amount of input. We illustrate how this works for group-theoretical fusion categories where the module categories have been classified \cite{ostrik2001module,10.1155/S1073792803205079,Natale_2017,etingof2019tensor} and provide the physical picture in terms of topological interfaces and their fusion.

We first recall that group-theoretical fusion categories are usually denoted as 
$\cC(G,\omega,H,\psi)$ where the defining data consists of a finite group $G$, its group 3-cocycle $\omega \in H^3(G,U(1))$, a subgroup $H\subset G$, and a 2-cochain $\psi \in C^2(H,U(1))$, subject to the condition that $d\psi=\omega|_H$. The special case $\cC(G,\omega,1,1)$ is the group category ${\rm \bf Vec}_G^\omega$ with a 't Hooft anomaly captured by the 3-cocycle $\omega$. The more general cases arise as dual categories under gauging algebra objects $A(H,\psi)$ in ${\rm \bf Vec}_G^\omega$, which are simply characterized by non-anomalous subgroups $H\subset G$, with a choice of the discrete torsion in $H^2(H,U(1))$ ($\psi$ is a torsor under this group). This coincides with the defining data of $\cC(G,\omega,H,\psi)$ up to auto-equivalence. The corresponding module category is denoted as $\cM(H,\psi)$, which is determined by $(H,\psi)$ up to conjugations in $(G,\omega)$ \cite{ostrik2001module,Natale_2017} and is the invertible bimodule category for the Morita equivalence between the two fusion categories. For example, ${\rm \bf Rep}(G)=\cC(G,1,G,1)$ is dual to ${\rm \bf Vec}_G$ by the module category $\cM(G,1)$, which has rank 1 (unique simple object), equivalently from the category of $(A(G,1),A(G,1))$-bimodules and $A(G,1)=\oplus_{g\in G} g$ is the regular representation of $G$. Such a rank 1 module category $\cM(G,1)\cong {\rm \bf Vec}$ realizes a fiber functor $F:\cC \to {\rm \bf Vec}={\rm Fun}({\rm \bf Vec},{\rm \bf Vec})$ (here for $\cC={\rm \bf Rep}(G)$) \cite{etingof2016tensor}. Physically, the module category $\cM(G,1)\cong {\rm \bf Vec}$ corresponds to the topological interface from imposing Dirichlet boundary conditions for the $G$-gauge fields in the half-gauging picture (see Section~\ref{sec:halfgauging}). The more general topological interfaces associated with the module category $\cM(H,\psi)$ correspond to partial Dirichlet (or mixed Neumann-Dirichlet) boundary conditions where only $H$-gauge fields are frozen at the interface.\footnote{The discrete torsion $\psi$ in $\cM(H,\psi)$ is introduced in the half-gauging picture for the $H$-gauge fields.}

\begin{figure}
    \centering
   \includegraphics[scale=1.6]{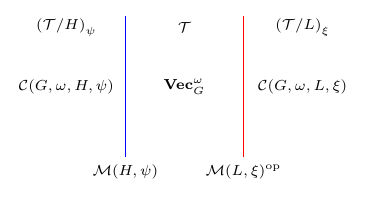}
    \caption{Topological interfaces for group theoretical fusion categories from invertible gauging and interface fusion.}
    \label{fig:modulecatsforgroup}
\end{figure}

As explained in Section~\ref{sec:modulecat}, (left) module categories over $\cC$ are simply the categories of topological interfaces on which TDLs in $\cC$ can end topologically from the left. Here we can construct such topological interfaces for $\cC(G,\omega,H,\psi)$ by starting with the topological interfaces for ${\rm \bf Vec}_G^\omega$ corresponding to the module category $\cM(L,\xi)$ with subgroup $L\subset G$ and 2-cochain $\xi$ satisfying $d\xi=\omega|_L$ and fusing with the topological gauging interfaces that relate the two or, equivalently, by gauging the non-anomalous subgroup with discrete torsion $(H,\psi)$ in a strip (see Figure~\ref{fig:modulecatsforgroup}). This gives an intuitive explanation for the classification of indecomposable module categories over $\cC(G,\omega,H,\psi)$ in \cite{Natale_2017,etingof2019tensor} by 
\ie 
\cM^{H,\psi}(L,\xi)={\rm Fun}_{{\rm \bf Vec}_G^\omega}(\cM(H,\psi),\cM(L,\xi))\,,
\label{gtmodulecat}
\fe
and $\cM^{H,\psi}(L,\xi)$ and $\cM^{H,\psi}(L',\xi')$ are equivalent if and only if $(L,\xi)$ and $(L',\xi')$ are related by conjugation in $(G,\omega)$.  
 
\subsection{Simple Dimensional Constraints for Generalized Gauging} 
\label{sec:dimconstraints}

While there are only limited results on the classification of (bi)module categories for fusion categories, there are a number of explicit and simple constraints on their quantum dimensions (both as a whole and for simple objects therein) from the quantum dimensions of the TDLs in the associated fusion categories. Such constraints are useful in identifying the possible generalized gaugings and symmetry properties of the gauged theories. Thus we review them below. All of these dimensional constraints have been proven for unitary fusion categories and can be found in the classic textbook \cite{etingof2016tensor}. Below, as we review them, we will also provide alternative arguments from basic axioms of QFT that realize such fusion category symmetries. It will become clear that all such dimensional constraints are simple consequences of unitarity (positivity) and locality of the underlying QFT, which is no surprise since the fusion categories and their (bi)module categories are just algebraic formulations of symmetries in QFT and the rigid structures of these categories are guaranteed by axioms of QFT.

\begin{figure}[!htb]
    \centering
    \includegraphics[scale=0.8]{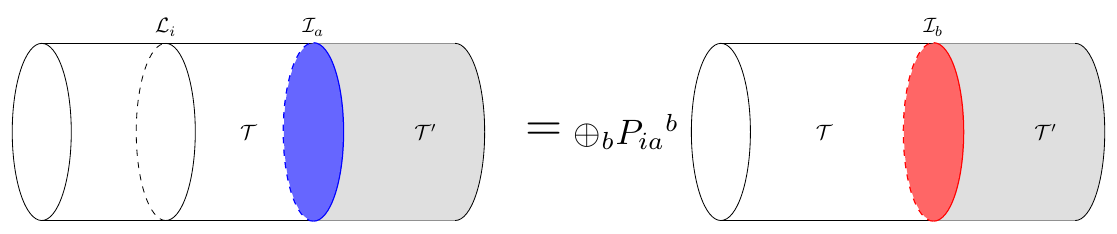}
    \caption{Dimension vector as the Frobenius-Perron eigenvector of the NIM-rep from fusion between TDL and interface.}
    \label{fig:dimensionvector}
\end{figure}

An important and elementary property of the NIM-rep $\{P_i\}$ for a left $\cC$-module category $\cM$ of rank $r$ is that the $r\times r$ matrices $P_i$ share a common \textit{positive} eigenvector $d_\cM$ known as the dimension vector of $\cM$ that contains the quantum dimensions of the $r$ simple topological interfaces $\cI_a\in \cM$ with $a=1,\dots,r$,
\ie 
P_i \cdot d_\cM =\la \cL_i\ra d_\cM\,,\quad d_\cM=\{\la \cI_a\ra\}\,.
\label{dimvector}
\fe 
This follows immediately from the defining equation for the NIM-rep \eqref{NIMrepdef} by taking the vacuum expectation value on the cylinder as in Figure~\ref{fig:dimensionvector}. The dimension vector $d_\cM$ is also the simultaneous Frobenius-Perron eigenvector of the non-negative matrices $P_i$ (the corresponding eigenvalue bounds from above the absolute value of all eigenvalues of $P_i$). It is a simple consequence of \eqref{dimvector} that the diagonal entries of $P_i$ are bounded by the quantum dimension of the TDL $\cL_i$,
\ie 
P_{ia}{}^a\leq \la \cL_i\ra\,,
\label{boundonP}
\fe
and the inequality is strict if $\la \cL_i\ra \notin \mZ_+$. We refer to the distinguished algebra $A$ when \eqref{boundonP} is saturated as the \textit{maximal} algebra (which may admit multiple inequivalent multiplication morphisms $m$),
\ie 
A_{\rm max}=\oplus_i \la \cL_i\ra \cL_i\,.
\label{amaxdef}
\fe
The maximal algebras in $\cC$ are in one-to-one correspondence with fiber functors $F:\cC\to {\rm\bf Vec}$, which exist if and only if the $\cC$ symmetry is anomaly-free in the definition of \cite{Thorngren:2019iar}. 

The general haploid algebra object associated with the interface $\cI_a$,
\ie 
A={\underline{\rm Hom}}(\cI_a,\cI_a)=\cI_a\otimes \overline{\cI}_a=\id \oplus_{i\neq \id}^{P_{ia}^a} \cL_i\,,
\label{Afrominterface}
\fe
is restricted by \eqref{boundonP} to a small number of possibilities. Since all haploid algebra objects arise this way \cite{ostrik2001module} (see also Section~\ref{sec:modulecat}), this reduces the classification of them to a finite problem (where the multiplication morphism $m$ has to be determined from the algebra conditions).

The total quantum dimension of a fusion category $\cC$ is commonly defined as
\ie 
\dim (\cC)\equiv \sum_i \la \cL_i\ra^2\,,
\fe
by the quantum dimensions of its simple TDLs $\cL_i$. We define the total quantum dimension of a $\cC$-module category $\cM$ in the analogous way,
\ie 
\dim (\cM)\equiv \sum_a \la \cI_a\ra^2\,,
\fe 
in terms of the simple interfaces $\cI_a\in \cM$. We then have the following equality 
\ie 
\dim (\cM) = \dim (\cC)\,,
\label{dimMC}
\fe
for any indecomposable $\cC$-module category $\cM$, which obviously bounds the rank of $\cM$ by
\ie 
{\rm rank} (\cM) \leq \dim (\cC)\,.
\fe
In fact we have the stronger equality
\ie 
R_\cM\equiv \sum_i \la \cL_i\ra P_{i}= d_\cM d_\cM^t\,, 
\label{dimmatrix}
\fe
from which \eqref{dimMC} follows by using \eqref{dimvector}. To derive \eqref{dimmatrix}, one first notes that 
\ie 
R_\cM^2= \dim(\cC) R_\cM\,,
\fe
using the NIM-rep and reciprocity properties of the fusion coefficients $N_{ij}^k$, which means the only nonzero eigenvalue of $R_\cM$ is $\dim(\cC)$. 
Since $R_\cM$ is a positive Hermitian matrix, the strongest version of the Frobenius-Perron theorem implies that $R_\cM$ has a unique eigenvector (up to overall normalization) with eigenvalue  $\dim(\cC)$, which is none other than $d_\cM$. Therefore $R_\cM$ is a rank-1 matrix proportional to $d_\cM d_\cM^t$, and the proportionality constant is fixed to 1 using \eqref{Afrominterface}.

Finally we note that the fusion category and its dual category under generalized gauging have identical total quantum dimensions \cite{etingof2016tensor},
\ie 
\dim (\cC)=\dim ({}_A \cC_A)\,.
\fe
In other words, categorical Morita equivalence preserves the total quantum dimension of the fusion category, and by \eqref{dimMC} also the total quantum dimension of its indecomposable module categories.\footnote{However the ranks of the indecomposable module categories may change under this 2-equivalence.} This follows from the fact that Morita equivalent categories $\cC$ and $\cC'$ have identical Drinfeld centers $Z(\cC)=Z(\cC')$ whose total quantum dimensions satisfy $\dim (Z(\cC))= \dim (\cC)^2$ \cite{etingof2016tensor}. We also note the following as a simple consequence of representing gauging by $A$ as topological interface $\cI_a$ where the algebra object is recovered by interface fusion as in $A=\cI_a\otimes \overline{\cI}_a$. The dual algebra object $A^*$ that undoes the gauging is simply $A^*=\overline{\cI}_a \otimes \cI_a$ from interface fusion in the opposite way, consequently $A$ and $A^*$ share the same quantum dimension,
\ie 
\la A\ra =\la A^*\ra =\la \cI_a\ra^2 \,.
\label{qdimAAdual}
\fe
This relation (together with \eqref{boundonP}) will become handy to identify the dual algebra object $A^*$ in explicit examples. 
For sequential gauging $\cC\to \cC' \to \cC''$ implemented by haploid algebras $A$ and $A'$ successively with corresponding simple topological interfaces $\cI$ and $\cI'$, if the topological interface between $\cC$ and $\cC''$ resulting from the fusion $\cI''=\cI\otimes \cI'$ remains simple,\footnote{This is not the case if $\cC''\cong \cC$ and $\cI'\cong \overline{\cI}$ which represents the topological interface that undoes the $A$ gauging by $A^*$, and \eqref{sequentialmultiplicative} does not apply.} the one-step gauging is simply implemented by the haploid algebra $A''=\cI''\otimes \overline{\cI''}$, which implies that, in particular,
\ie 
\la A''\ra = \la A\ra \la A'\ra\,,
\label{sequentialmultiplicative}
\fe
generalizing the familiar result for invertible sequential gauging.

\section{Generalized Gauging in ${\rm \bf Rep}(H_8)$ and ${\rm \bf Rep}(D_8)$}
\label{sec:gengaugingRep}

To illustrate the general features of generalized gauging discussed in Section~\ref{sec:genprop}, we study two examples of fusion categories ${\rm \bf Rep}(H_8)$ and ${\rm \bf Rep}(D_8)$ here in detail, which admit a variety of non-invertible TDLs that can be gauged. Furthermore these fusion category symmetries are realized in large families of CFTs, and as we will see in Section~\ref{sec:CFTrealization}, so are their extensive structures under generalized gauging. 

We first recall that ${\rm \bf Rep}(H_8)$ and ${\rm \bf Rep}(D_8)$ are two inequivalent fusion categories of rank 5 with simple TDLs $\{\id,\eta,\eta',\eta\eta',\cV\}$ such that $\eta,\eta'$ are invertible $\mZ_2$ TDLs that generate a ${\rm \bf Vec}_{\mZ_2\times \mZ_2}$ subcategory, and the self-dual $\cV$ provides a non-invertible $\mZ_2$-extension with the following fusion rules,
\ie 
\cV\otimes \cV=\id \oplus  \eta \oplus\eta'\oplus\eta\eta'\,,\quad \cV \eta =\eta \cV=\cV\eta'=\eta'\cV=\cV\,,
\fe
which generates the Tambara-Yamagami (TY) fusion ring ${\rm \bf TY}(G)$  for abelian group $G=\mZ_2\times \mZ_2$ \cite{tambara1998tensor}. The non-invertible TDL $\cV$ is often referred to as the duality defect and the TY symmetry as the duality symmetry because of the relation to Kramers-Wannier duality in $d=2$ \cite{Frohlich:2004ef}. 

Each TY fusion ring ${\rm \bf TY}(G)$ admits a finite set of inequivalent F-symbols that define the corresponding fusion category \cite{tambara1998tensor}. 
Here there are four possibilities ${\rm \bf TY}(\mZ_2^2,\chi,\ep)$ determined by a symmetric non-degenerate bicharacter $\chi=\chi_{s,a}$ on $\mZ_2^2$,
\ie 
\chi_s(i_1,i_2;j_1,j_2)=(-1)^{i_1 j_1+i_2j_2}\,,\quad \chi_a(i_1,i_2;j_1,j_2)=(-1)^{i_1 j_2+i_2j_1}\,,
\fe
for $i_1,i_2,j_1,j_2\in \mZ_2$ and the Frobenius-Schur indicator $\ep=\pm 1$ for the duality defect $\cV$. In this notation, the ${\rm \bf Rep}(H_8)$ and ${\rm \bf Rep}(D_8)$ fusion categories are
\ie 
{\rm \bf Rep}(H_8)={\rm \bf TY}(\mZ_2^2,\chi_s,+)\,,\quad {\rm \bf Rep}(D_8)={\rm \bf TY}(\mZ_2^2,\chi_a,+)\,,
\fe
which makes clear that the corresponding TY category admits a fiber functor and can be realized as the representation category of a finite group or more generally a Hopf algebra. Here $D_8$ is the dihedral group of order 8 and $H_8$ is the unique dimension 8 Hopf algebra of Kac and Paljutkin that is neither commutative nor cocommutative \cite{kac1966finite}.\footnote{The Kac-Paljutkin Hopf algebra $H_8$ is also the lowest dimensional semisimple Hopf algebra that is not the group algebra $\mC[G]$ of a finite group $G$. It has generators $x,y,z$ which satisfy the following algebra relations
\ie 
x^2=y^2=1\,,~z^2={1\over 2}(1+x+y-xy)\,,~zx=yz\,,~zy=xz\,,~xy=yx\,,
\label{H8alg}
\fe 
which obviously contains a $\mZ_2\times \mZ_2$ group subalgebra.
}

The nontrivial F-symbols for ${\rm \bf Rep}(H_8)$ and ${\rm \bf Rep}(D_8)$ in the \textit{positive} gauge (see Figure~\ref{fig:Fpositivegauge}) take the following form
\ie 
[F_{\cV\cV\cV}^\cV]_{g,h}={1\over 2}\chi(g,h)^{-1}\,,\quad 
[F_{\cV g \cV}^h]_{\cV,\cV}=[F_{g\cV h}^\cV]_{\cV,\cV}=\chi(g,h)\,,
\label{TYFsymbol}
\fe
where $\chi=\chi_{s,a}$ and all other components of the F-symbols are 1 (if the fusion channel exists) or 0.

Finally, both ${\rm \bf Rep}(H_8)$ and ${\rm \bf Rep}(D_8)$ are also group-theoretical fusion categories,\footnote{We emphasize that one given group-theoretical fusion category may have multiple representations $\cC(G,\omega,H,\psi)$ that differ in the group-theoretical data.}
\ie 
{\rm \bf Rep}(H_8)=\cC(D_8,\C,\mZ_2^s,1)\,,\quad {\rm \bf Rep}(D_8)=\cC(D_8,1,D_8,1)\,,
\label{H8D8grouptheoretical}
\fe
where $\mZ_2^s\subset D_8$ is the non-anomalous non-normal subgroup $\la s\ra$ in the following standard representation of $D_8$,
\ie 
D_8=\la r,s|r^4=s^2=1\,,~sr=r^{-1}s\ra \,,
\label{D8}
\fe
and $\C\in H^3(D_8,U(1)) \cong \mZ_4\times \mZ_2^2$ is an order-two 3-cocycle taking values in $\{\pm 1\}\subset U(1)$
as follows: $\C(s^{m_1}r^{n_1}, s^{m_2}r^{n_2}, s^{m_3}r^{n_3})=1$ unless $m_1$ is odd and we are in one of the following three cases: $n_2 = 1$ and $s^{m_3}r^{n_3} \in \{r^2,sr^3\}$; $n_2 = 2$ and $s^{m_3}r^{n_3} \in \{r,r^2,sr^3,s\}$; $n_2 = 3$ and $s^{m_3}r^{n_3} \in \{r^3,sr^2\}$.\footnote{We have explicitly checked that this is desired cocycle using GAP \cite{GAP4} and SageMath \cite{SageMath} and applying the universal coefficient theorem for group cohomology.}

Consequently the module categories of ${\rm \bf Rep}(H_8)$ and ${\rm \bf Rep}(D_8)$ have been classified by methods in \cite{Natale_2017,etingof2019tensor} specialized for group-theoretical categories. In the following we will re-derive these results in a different way that has the potential to apply in much greater generality \cite{DLWW2}. We also present explicit expressions for the algebra objects that were not available previously. Finally we will determine the Brauer-Picard groupoids for these fusion categories that capture identifications between different sequential gaugings.

\subsection{NIM-reps for ${\rm \bf TY}(\mZ_2^2)$ Fusion Ring}
\label{sec:NIMrepsTY}

As a preparation towards fully classifying the algebra objects (and the corresponding module categories), we first study the possible NIM-reps, which capture the first layer of data in the module categories, using only the data of the fusion ring of the underlying fusion category. Even without putting in the information of the F-symbols this is already quite constraining, as is emphasized in \cite{grossman2012quantum,grossmansnyderAH} which studies the algebra objects and the Brauer-Picard groupoid for the Haagerup fusion category and its generalizations.

We adopt the following strategy that improves upon the algorithm presented in \cite{grossman2012quantum,grossmansnyderAH}. The details and general applications will be addressed in \cite{DLWW2}; here we will briefly summarize the main ideas in the physical picture using the topological interfaces introduced in Section~\ref{sec:genprop}. For the moment we keep the underlying fusion category $\cC$ (and its fusion ring) general and will later specialize to the ${\rm \bf TY}(\mZ_2^2)$ case that is relevant here.

We first enumerate the possible non-simple TDLs $A=\id \oplus_{i\neq \id}^{n_i} \cL_i \in \cC$ subject to the condition $n_i\leq \la \cL_i\ra$ from \eqref{boundonP}. We then study the interface fusion equation \eqref{Afrominterfacefusion} on the cylinder, sandwiched between simple TDLs $\cL_i,\cL_j\in \cC$ (here for ${\rm \bf Rep}(H_8)$)
\ie 
  \cL_i \otimes A\otimes \overline{\cL}_j  =\cL_i \otimes (\cI_a\otimes \overline{\cI}_a)\otimes \overline{\cL}_j  \,,
\fe 
where it's important to remember that the algebra $A$ may depend on the choice of $\cI_a$ (however, the Morita equivalence class of $A$ is fixed). Projecting onto the ground state on the circle (e.g. taking the cylinder to be thin), we have
\ie 
M_{ij} \equiv \dim {\rm Hom}(\cL_i\otimes A,\cL_j) = \sum_b P_{ia}{}^b P_{jb}{}^a \,,
\fe
where the matrix $M_{ij}$ is easily determined from the fusion ring and the candidate $A$.\footnote{Using the Frobenius recipricocity of the fusion ring one can show such an $M_{ij}$ exists for any NIM-rep regardless of whether or not it corresponds to a module category. Therefore, the algorithm detailed here will find all NIM-reps of a given fusion ring.} We then perform matrix factorization to find possible NIM-rep matrices $P_i$ satisfying \eqref{NIMrepdef} of dimension $r$ subject to the dimensional constraints in Section~\ref{sec:dimconstraints}, which also further restricts $A$. Each NIM-rep $\cM=\{P_i\}$ derived this way has a corresponding dimension vector $d_\cM$ from the Frobenius-Perron eigenvector (see Section~\ref{sec:dimconstraints}). 
Two NIM-reps $\cM=\{P_i\}$ and $\cM'=\{P'_i\}$ of the same dimension $r$ are equivalent if they are related by an
$S_r$ permutation of $\{\cI_a\}$ (in particular $d_\cM=d_{\cM'}$ up to permutation). 
The above procedure produces a collection of irreducible and inequivalent NIM-reps which are candidates for full-fledged module categories for the fusion categories that share the input fusion ring.
For each of these NIM-irreps, the corresponding algebra objects can be read-off from the diagonal entries of $P_i$ as in \eqref{Afrominterface}.

Before we present the resulting list of (irreducible and inequivalent) NIM-reps for ${\rm \bf TY}(\mZ_2^2)$ we first note that 
this fusion ring has a nontrivial ``triality'' outer-automorphism 
\ie 
{\rm Out}({\rm \bf TY}(\mZ_2^2))=S_3\,,
\label{outfusionring}
\fe
that permutes the three invertible TDLs. Consequently, the NIM-reps, in general, come in families related by this $S_3$ automorphism which, in general, does not act faithfully on the NIM-reps due to the equivalence relation by permuting the basis elements. 

Starting with the highest rank NIM-irrep, we first have a unique rank 5 NIM-irrep that coincides with the regular NIM-irrep that exists from any fusion ring,
\ie
{\rm NR}_1=&\left\{\left(\begin{array}{rrrrr}
1 & 0 & 0 & 0 & 0 \\
0 & 1 & 0 & 0 & 0 \\
0 & 0 & 1 & 0 & 0 \\
0 & 0 & 0 & 1 & 0 \\
0 & 0 & 0 & 0 & 1
\end{array}\right)\,, \left(\begin{array}{rrrrr}
0 & 1 & 0 & 0 & 0 \\
1 & 0 & 0 & 0 & 0 \\
0 & 0 & 0 & 1 & 0 \\
0 & 0 & 1 & 0 & 0 \\
0 & 0 & 0 & 0 & 1
\end{array}\right)\,, \left(\begin{array}{rrrrr}
0 & 0 & 1 & 0 & 0 \\
0 & 0 & 0 & 1 & 0 \\
1 & 0 & 0 & 0 & 0 \\
0 & 1 & 0 & 0 & 0 \\
0 & 0 & 0 & 0 & 1
\end{array}\right)\,,\right.
\\
&\left. \left(\begin{array}{rrrrr}
0 & 0 & 0 & 1 & 0 \\
0 & 0 & 1 & 0 & 0 \\
0 & 1 & 0 & 0 & 0 \\
1 & 0 & 0 & 0 & 0 \\
0 & 0 & 0 & 0 & 1
\end{array}\right)\,, \left(\begin{array}{rrrrr}
0 & 0 & 0 & 0 & 1 \\
0 & 0 & 0 & 0 & 1 \\
0 & 0 & 0 & 0 & 1 \\
0 & 0 & 0 & 0 & 1 \\
1 & 1 & 1 & 1 & 0
\end{array}\right) \right\}\,,\quad 
A_1= \,\id \,,\id \oplus \eta \oplus \eta'\oplus \eta\eta'\,,
\label{NR1}
\fe
and we have also listed the corresponding haploid algebra objects which are Morita equivalent.

We next have a family of $3 = \binom{3}{2}$ rank 4 NIM-irreps related by the $S_3$ automorphism to the following,
\ie
{\rm NR}_2=&\left\{\left(\begin{array}{rrrr}
1 & 0 & 0 & 0 \\
0 & 1 & 0 & 0 \\
0 & 0 & 1 & 0 \\
0 & 0 & 0 & 1
\end{array}\right)\,, \left(\begin{array}{rrrr}
0 & 1 & 0 & 0 \\
1 & 0 & 0 & 0 \\
0 & 0 & 1 & 0 \\
0 & 0 & 0 & 1
\end{array}\right)\,,  \left(\begin{array}{rrrr}
0 & 1 & 0 & 0 \\
1 & 0 & 0 & 0 \\
0 & 0 & 0 & 1 \\
0 & 0 & 1 & 0
\end{array}\right)\,, \left(\begin{array}{rrrr}
1 & 0 & 0 & 0 \\
0 & 1 & 0 & 0 \\
0 & 0 & 0 & 1 \\
0 & 0 & 1 & 0
\end{array}\right)\,, \right.
\\
&\left.\left(\begin{array}{rrrr}
0 & 0 & 1 & 1 \\
0 & 0 & 1 & 1 \\
1 & 1 & 0 & 0 \\
1 & 1 & 0 & 0
\end{array}\right)\right\}\,,\quad 
 A_2= \id \oplus \eta\eta'\,, \id \oplus \eta\,.
 \label{NR2}
\fe
There is another family of $3 = \binom{3}{2}$ rank 4 NIM-irreps related by $S_3$ to,
\ie
{\rm NR}_3=&\left\{\left(\begin{array}{rrrr}
1 & 0 & 0 & 0 \\
0 & 1 & 0 & 0 \\
0 & 0 & 1 & 0 \\
0 & 0 & 0 & 1
\end{array}\right)\,, \left(\begin{array}{rrrr}
0 & 1 & 0 & 0 \\
1 & 0 & 0 & 0 \\
0 & 0 & 0 & 1 \\
0 & 0 & 1 & 0
\end{array}\right)\,, \left(\begin{array}{rrrr}
0 & 1 & 0 & 0 \\
1 & 0 & 0 & 0 \\
0 & 0 & 0 & 1 \\
0 & 0 & 1 & 0
\end{array}\right)\,, \left(\begin{array}{rrrr}
1 & 0 & 0 & 0 \\
0 & 1 & 0 & 0 \\
0 & 0 & 1 & 0 \\
0 & 0 & 0 & 1
\end{array}\right)\,, \right.
\\
&\left.\left(\begin{array}{rrrr}
0 & 0 & 1 & 1 \\
0 & 0 & 1 & 1 \\
1 & 1 & 0 & 0 \\
1 & 1 & 0 & 0
\end{array}\right)\right\}\,,\quad 
A_3= \id \oplus \eta\eta'\,,
\label{NR3}
\fe
There is no NIM-irrep of rank 3. At rank 2, there are $3 = \binom{3}{2}$ NIM-irreps related by $S_3$ to,
\ie
{\rm NR_4}=\left\{\left(\begin{array}{rr}
1 & 0 \\
0 & 1
\end{array}\right)\,, \left(\begin{array}{rr}
0 & 1 \\
1 & 0
\end{array}\right)\,, \left(\begin{array}{rr}
0 & 1 \\
1 & 0
\end{array}\right)\,, \left(\begin{array}{rr}
1 & 0 \\
0 & 1
\end{array}\right)\,, \left(\begin{array}{rr}
1 & 1 \\
1 & 1
\end{array}\right)\right\}\,,~ A_4= \id \oplus \eta\eta' \oplus \cV\,,
\label{NR4}
\fe
and another rank 2 NIM-irrep that is $S_3$-invariant,
\ie
{\rm NR_5}=\left\{\left(\begin{array}{rr}
1 & 0 \\
0 & 1
\end{array}\right)\,, \left(\begin{array}{rr}
1 & 0 \\
0 & 1
\end{array}\right)\,, \left(\begin{array}{rr}
1 & 0 \\
0 & 1
\end{array}\right)\,, \left(\begin{array}{rr}
1 & 0 \\
0 & 1
\end{array}\right)\,, \left(\begin{array}{rr}
0 & 2 \\
2 & 0
\end{array}\right)\right\}\,,~ A_5= \id \oplus \eta\oplus \eta' \oplus \eta\eta'\,.
\label{NR5}
\fe
Finally, there is a unique rank 1 NIM-irrep which is again $S_3$-invariant,
\ie
{\rm NR}_6=\left\{\left(\begin{array}{r}
1
\end{array}\right)\,, \left(\begin{array}{r}
1
\end{array}\right)\,, \left(\begin{array}{r}
1
\end{array}\right)\,, \left(\begin{array}{r}
1
\end{array}\right)\,, \left(\begin{array}{r}
2
\end{array}\right)\right\}\,,\quad A_6= \id \oplus \eta \oplus  \eta'\oplus  \eta\eta' \oplus 2\cV\,.
\label{NR6}
\fe

\subsection{Algebra objects in ${\rm \bf Rep}(H_8)$}
\label{sec:RepH8alg}

In the previous section, we have seen that consistency of interface fusion produces a small list of candidate algebra objects (and corresponding module categories). We now finish the classification by solving the algebra conditions for the candidate $A$ objects using the explicit F-symbols. We emphasize that, in general, there could be multiple multiplication morphisms $m$ for a given $A$ as a non-simple TDL in $\cC$.

For $\cC={\rm \bf Rep}(H_8)$ with F-symbols given in \eqref{TYFsymbol}, carrying out the above procedure, we find 8 different algebra objects which are listed in Table~\ref{tab:repH8}. They fall into 6 Morita equivalence classes where the Morita trivial class includes two algebra objects given by $A=\id$ and $A=\cV\otimes \cV$ (see around Theorem~\ref{thm:selfdual}), a Morita non-trivial class includes $A=\id \oplus \eta$ and $A=\id \oplus \eta'$, and the remaining Morita classes have unique algebra objects. Most of these algebra objects have an invertible origin, namely, they are associated with gauging subgroups of the non-anomalous $\mZ_2\times \mZ_2$ subcategory with discrete torsion; the corresponding multiplication morphisms are listed below,
\ie
A=\id\oplus\eta:\quad m_{11}^1 = m_{1\eta}^{\eta} = m_{\eta 1}^{\eta} = m_{\eta \eta}^1 = \frac{1}{\sqrt{2}}\,,
\label{Z2mmap}
\fe
and similarly for $A=\id\oplus\eta'$ and $A=\id\oplus\eta\eta'$ by relabeling the legs of $m^i_{jk}$. 
For $\mZ_2\times \mZ_2$ gauging, there are two choices of discrete torsion, 
\ie
\label{Z2Z2mmap}
A=\id\oplus\eta\oplus\eta'\oplus\eta\eta':\,\quad m_{g,h}^{gh} = \frac{1}{2} \begin{pmatrix}
    1 & 1 & 1 & 1 \\
     1 & 1 & 1 & 1 \\
      1 & 1 & 1 & 1 \\
       1 & 1 & 1 & 1
\end{pmatrix}
\,,~ \frac{1}{2} \begin{pmatrix}
    1 & 1 & 1 & 1 \\
     1 & 1 & i & -i \\
      1 & -i & 1 & i \\
       1 & i & -i & 1
\end{pmatrix}\,,
\fe
where $g,h\in \mZ_2\times \mZ_2$. We differentiate the cases by the $\star$ subscript in Table~\ref{tab:repH8}.
Finally, 
we highlight the two non-invertible gaugings given by 
\ie 
A^{\rm I}\equiv \id \oplus \eta\eta'\oplus \cV\,,\quad A^{\rm II}\equiv \id \oplus \eta\oplus \eta'\oplus \eta\eta'\oplus 2\cV\,,
\fe 
with the following unique
 multiplication morphisms
\ie
A=A^{\rm I}:\quad 
&m_{11}^1 = m_{1,\eta\eta'}^{\eta\eta'} = m_{\eta\eta', 1}^{\eta\eta'} = m_{\eta\eta', \eta\eta'}^1 = m_{1\cV}^{\cV} = m_{\cV 1}^{\cV} = m_{\cV\cV}^1 \\
&= m_{\eta\eta',\cV}^{\cV} = m_{\cV,\eta\eta'}^{\cV} 
= m_{\cV\cV}^{\eta\eta'} = \frac{1}{2}\,,
\fe
and finally the unique maximal algebra object (see \eqref{amaxdef} for definition)
\ie
A=A^{\rm II}:\quad & m_{g,h}^{gh} = \frac{1}{2\sqrt{2}} \begin{pmatrix}
    1 & 1 & 1 & 1 \\
     1 & 1 & -i & i \\
      1 & i & 1 & -i \\
       1 & -i & i & 1
\end{pmatrix} \,, m_{1\cV}^{\cV} = m_{\cV 1}^{\cV} = m_{\cV\cV}^1 = \frac{1}{2\sqrt{2}}\begin{pmatrix}
    1 & 0 \\
    0 & 1
\end{pmatrix} \,, \\
&m_{\eta\eta',\cV}^{\cV} = m_{\cV,\eta\eta'}^{\cV} = m_{\cV,\cV}^{\eta\eta'} = \frac{1}{2\sqrt{2}} \begin{pmatrix}
    1 & 0 \\
    0 & -1
\end{pmatrix} \,, \\
&m_{\eta,\cV\alpha}^{\cV\beta} = m_{\eta',\cV\beta}^{\cV\alpha} = m_{\cV\beta,\eta}^{\cV\alpha} = m_{\cV\alpha,\eta'}^{\cV\beta} = m_{\cV\alpha,\cV\beta}^{\eta} = m_{\cV\beta,\cV\alpha}^{\eta'} = \frac{1}{2\sqrt{2}} \begin{pmatrix}
    0 & -e^{\frac{\pi i}{4}} \\
    e^{\frac{3\pi i}{4}} & 0
\end{pmatrix}_{\alpha\beta}\,.
\fe

\begin{table}[!htb]
    \centering
    \begin{tabular}{|c|c|c|c|}
    \hline
   $L\subset D_8$ &    Algebra object $A$   & Module Category $\cC_A$ & Dual Fusion Category ${}_A \cC_A$ \\\hline  \hline
$\la s\ra $     & $\id,\id\oplus\eta\oplus\eta'\oplus\eta\eta'$  & ${\rm NR}_1$: ${\rm \bf Rep}(H_8)$ & ${\rm \bf Rep}(H_8)$
     \\\hline 
   $\la s,r^2\ra $    &    $\id\oplus\eta,\id \oplus\eta'$  
    & ${\rm NR}_2$: $\{A,\eta'\oplus\eta'\eta, \cV_1,\cV_2\}$ &  ${\rm \bf Rep}({H_8})$
         \\\hline 
   1   &    $\id \oplus\eta \eta'$  
    & ${\rm NR}_3$: $\{A,\eta\oplus \eta', \cV_1,\cV_2\}$ & ${\rm \bf Vec}_{D_8}^\C$
      \\\hline 
   $\la r^2\ra $     &   $(\id\oplus\eta\oplus\eta'\oplus\eta\eta')_\star$  
    &${\rm NR}_5$: $\{A,2\cV\}$ & ${\rm \bf Vec}_{D_8}^\C$
     \\\hline 
    $\la sr\ra $    &   ${\id\oplus\eta\eta'\oplus \cV}$  
    & ${\rm NR}_4$: $\{A,\eta\oplus \eta'\oplus \cV\}$ & ${\rm \bf Rep}(H_8)$
      \\\hline 
   $\la sr,r^2\ra $    &    ${\id\oplus \eta\oplus\eta'\oplus\eta\eta'\oplus 2\cV}$  
    & ${\rm NR}_6$: $\{A\}$ & ${\rm \bf Rep}(H_8)$ \\\hline
    \end{tabular}
    \caption{Algebra objects $A$ in ${\rm \bf Rep}(H_8)$ grouped according to the Morita equivalence classes together with the corresponding module categories $\cC_A$ (simple objects therein and the corresponding NIM-rep from Section~\ref{sec:NIMrepsTY}) and dual fusion categories ${}_A\cC_A$. The $\star$ subscript denotes nontrivial discrete torsion for the corresponding algebra and module category. The first column labels the module categories using group-theoretical data in $D_8$ according to \eqref{gtmodulecat} for $(H,\psi)=(\la s\ra,1)$.}
    \label{tab:repH8}
\end{table}

\subsection{Module Categories and the Orbifold Groupoid for ${\rm \bf Rep}(H_8)$}
\label{sec:H8modulecat}

Having classified the algebra objects $(A,m)$ in ${\rm \bf Rep}(H_8)$, the corresponding NIM-irreps immediately follow from Section~\ref{sec:NIMrepsTY} and are listed in Table~\ref{tab:repH8}. Note that the $S_3$ automorphism of the TY fusion ring \eqref{outfusionring} does not preserve the F-symbols \eqref{TYFsymbol} for ${\rm \bf Rep}(H_8)$ except for the $\mZ_2$ subgroup that exchanges $\eta$ and $\eta'$ \cite{tambara2000representations},
\ie 
{\rm Out}({\rm \bf Rep}(H_8))=\mZ_2\,.
\label{outRepH8}
\fe

\paragraph{Module categories for ${\rm \bf Rep}(H_8)$}
The rank of each indecomposable module category $\cM$ follows from that of the corresponding NIM-irrep, and in the third column of Table~\ref{tab:repH8} we list the simple objects in $\cM$ as simple right $A$-modules (written as non-simple TDLs in $\cC$) using the equivalence $\cM\cong \cC_A$ \cite{ostrik2001module}. Note that all simple $A$-modules can be determined from (submodules of) induced $A$-modules (namely $A\otimes \cL_i$ for simple TDLs $\cL_i\in \cC$) in a standard procedure \cite{Fuchs:2002cm}, by taking into account possible nontrivial endomorphisms of the induced module that dictates how such an induced module splits into simple $A$-modules.\footnote{This is the direct generalization of the usual fixed point resolution in ordinary orbifold \cite{Schellekens:1990xy} for both local operators and defects.} Here this splitting happens for the two rank 4 module categories, where the induced module $A\otimes \cV$ decomposes into two simple $A$-modules that are differentiated by the subscript on $\cV$, and the corresponding NIM-irreps are given in \eqref{NR2} and \eqref{NR3}. For the algebra object $A=\id\oplus\eta\oplus\eta'\oplus\eta\eta'$ there are two inequivalent multiplication morphisms that differ by the discrete torsion $H^2(\mZ_2^2,U(1))=\mZ_2$ (differentiated by the subscript $\star$ in Table~\ref{tab:repH8}). One of them is Morita equivalent to trivial gauging where the multiplication morphism is the trivial canonical one for $A=\cV\otimes \cV$, and the induced module $A\otimes \cV$ splits into four simple $A$-modules. The resulting module category is the regular module category that is universal to any fusion category with the corresponding NIM-irrep \eqref{NR1}.
In the other case, this splitting is forbidden due to the nontrivial multiplication morphism, and the resulting module category is rank 2 and corresponds to the NIM-irrep  \eqref{NR5}. Finally the two non-invertible gaugings where $\cV\in A$ produce a rank 2 module category and a rank 1 module category with NIM-irreps \eqref{NR4} and \eqref{NR6}, respectively. The rank 1 module category ${\rm \bf Vec}$ with the unique maximal algebra object realizes the unique fiber functor for ${\rm \bf Rep}(H_8)$  \cite{tambara2000representations}. 

Finally these module categories can also be derived using group-theoretical methods and are denoted as $\cM^{\mZ_2^s,1}(L,\xi)$ for a subgroup $L\subset D_8$ and 2-cochain $\xi$ satisfying $d\xi=\C|_L$ (up to conjugations in $D_8$) in \eqref{gtmodulecat} \cite{etingof2019tensor}. The nontrivial 3-cocycle $\C\in H^3(D_8,U(1))$ of order 2 is such that the following subgroups of $D_8$ (up to conjugacy) are non-anomalous (i.e. $\C$ is trivialized in $L$),
\ie 
L=1\,,~\la s \ra\,,~\la sr \ra\,,~\la r^2\ra\,,~\la s,r^2 \ra\,,~\la sr,r^2 \ra\,.
\label{Llist}
\fe
The last two subgroups in the above list can support a nontrivial discrete torsion (SPT) from
$H^2(\mZ_2\times \mZ_2,U(1))=\mZ_2$. However, the mixed anomaly (due to the 3-cocycle $\C$) between $\la s\ra$ and $\la sr,r^2\ra$ (similarly $\la sr\ra$ and $\la s,r^2\ra$) means that the SPT phase can be absorbed by implementing an $s$ symmetry transformation, which is a $d=2$ discrete version of the famous chiral Adler-Bell-Jackiw (ABJ) anomaly. Consequently, adding the discrete torsion to subgroups in \eqref{Llist} does not produce physically distinct gaugings in ${\rm \bf Vec}_{D_8}^\C$, and the inequivalent and indecomposable module categories are simply $M^{\mZ_2^s,1}(L,1)$ for $L$ in \eqref{Llist}. They are in one-to-one correspondence with the six module categories for ${\rm \bf Rep}(H_8)$ we have found as tabulated in Table~\ref{tab:repH8} \cite{etingof2019tensor}.

\paragraph{Dual fusion categories for ${\rm \bf Rep}(H_8)$}

Having determined the module categories we move on to discuss the dual fusion categories ${}_A \cC_A$, which capture a universal part of the symmetries of the theory after gauging $A$. The dual fusion category is equivalent to the category of $(A,A)$-bimodules \cite{ostrik2001module}, and the latter can be worked out in a similar way, although more tedious, as for the $A$-modules, using induced bimodules \cite{Fuchs:2002cm}. Here, instead, we will argue based on the general structure of generalized gauging and consistency conditions thereof (see Section~\ref{sec:genprop}, in particular, Section~\ref{sec:dimconstraints}), that the dual categories are given by those in the last column of Table~\ref{tab:repH8}.

We start with the realization of ${\rm \bf Rep}(H_8)$ as a group-theoretical category given in \eqref{H8D8grouptheoretical}, which says ${\rm \bf Rep}(H_8)$ is the dual category to ${\rm \bf Vec}_{D_8}^\C$ under gauging the $\mZ_2^s$ subgroup, or equivalently in terms of the notation mentioned at the end of Section~\ref{sec:modulecat},
\ie 
{\rm \bf Rep}(H_8) = (({\rm \bf Vec}_{D_8}^\C)_{\cM(\mZ_2^s,1)}^*)^{\rm op}\,,
\fe 
where the module category $\cM(\mZ_2^s,1)$ over ${\rm \bf Vec}_{D_8}^\C$ defines the category of topological interfaces that interpolate between the two dual fusion categories. Conversely, as is the case for any invertible gauging of a finite abelian group $G$, there is a dual invertible symmetry $\hat G\cong G$ (known as quantum symmetry \cite{Vafa:1989ih} in orbifold models), that can be gauged to undo the $G$-gauging. Here this means there must be a gaugeable $\mZ_2$ subcategory in ${\rm \bf Rep}(H_8)$ that implements the reverse gauging,
\ie 
{\rm \bf Vec}_{D_8}^\C = (({\rm \bf Rep}(H_8)_{\cM(\hat\mZ_2,1)}^*)^{\rm op}\,,
\fe
where the ${\rm \bf Rep}(H_8)$-module category $\cM(\hat\mZ_2,1)$ is equivalent to the opposite of the ${\rm \bf Vec}_{D_8}^\C$-module category $\cM(\mZ_2^s,1)$ as an invertible bimodule category over $({\rm \bf Rep}(H_8),{\rm \bf Vec}_{D_8}^\C)$. All $\mZ_2$ symmetries in ${\rm \bf Rep}(H_8)$ are clearly gaugeable, however the special one $\hat\mZ_2$ needed here must be the subgroup generated by the invertible TDL $\eta\eta'$ which is distinguished by having trivial F-symbols (i.e. 1 if the fusion channel exists) with all TDLs in ${\rm \bf Rep}(H_8)$. This ensures that the duality defect $\cV$ under this $\hat\mZ_2$ gauging can split consistently into a pair of conjugate invertible TDLs that correspond to generators of the $\mZ_4$ subgroup of dual $D_8$ symmetry. We thus identify ${\rm \bf Vec}_{D_8}^\C$ as the dual category for gauging $A=\id \oplus \eta\eta'$ in Table~\ref{tab:repH8}. 

The rest of the list of dual fusion categories then follows by sequential gauging starting (see Section~\ref{sec:seqgauginggrouptheoretical}) from ${\rm \bf Vec}_{D_8}^\C$ by gauging $L\subset D_8$ in \eqref{Llist}. For example, gauging the non-normal subgroups $L=\la s\ra\,,\la sr\ra$ again gives ${\rm \bf Rep}(H_8)$ while gauging the center $L=\la r^2\ra$ produces ${\rm \bf Vec}_{D_8}^\C$, which is a consequence of the nontrivial extension,
\ie 
1 \to \mZ_2^{r^2} \to D_8 \to \mZ_2\times \mZ_2 \to 1
\label{D8extension}
\fe
characterized by a nontrivial class of $H^2(\mZ_2\times \mZ_2,\mZ_2)=\mZ_2^3$ in the presence of a nontrivial mixed anomaly $\C$ \cite{Tachikawa:2017gyf}. 

We finish this analysis of dual symmetries under gauging algebra objects by remarking that the dual category of the representation category ${\rm \bf Rep}(H)$ of a general semisimple Hopf algebra $H$ with respect to a fiber functor (i.e. for module category ${\rm \bf Vec}$ and e.g. the forgetful functor) is given by ${\rm \bf Rep}(H^*)^{\rm op}$ for the dual Hopf algebra $H^*$ \cite{etingof2016tensor}. The dimension 8  Kac-Paljutkin algebra $H_8$ is self-dual $H_8\cong H_8^*$, which follows from the uniqueness property mentioned around \eqref{H8alg}. Consequently, the dual category under gauging the maximal algebra object in ${\rm \bf Rep}(H_8)$ is equivalent to ${\rm \bf Rep}(H_8)$ itself.\footnote{Here we have used that ${\rm \bf Rep}(H_8)\cong {\rm \bf Rep}(H_8)^{\rm op}$, which is obvious from its fusion rules and F-symbols.}
Of course this is consistent with the analysis above based on sequential gauging from ${\rm \bf Vec}_{D_8}^\C$.

\begin{figure}[!htb]
    \centering
    \includegraphics[scale=2.5]{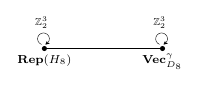}
    \caption{Generalized orbifold groupoid for ${\rm \bf Rep}(H_8)$.}
    \label{fig:H8groupoid}
\end{figure}

\paragraph{Generalized orbifold groupoid for ${\rm \bf Rep}(H_8)$}
Now that we have completed explicitly specifying all algebraic properties of generalized gauging in ${\rm \bf Rep}(H_8)$, we are ready to present the corresponding generalized orbifold groupoid (Brauer-Picard groupoid) that neatly packages the Morita equivalences between categorically Morita equivalent fusion categories: namely, the dual symmetries that arise under generalized gauging (see Figure~\ref{fig:H8groupoid}). As explained in Section~\ref{sec:seqgauginggrouptheoretical} for the general case of a connected subgroupoid denoted as ${\rm \bf BrPic}$ containing fusion category $\cC$, such Morita equivalences are in one-to-one correspondence with invertible bimodules over a pair of fusion categories dual to $\cC$, and these invertible bimodules all arise from indecomposable module categories. In particular, the $(\cC,\cC)$-bimodule categories generate the group of automorphisms, denoted as ${\rm \bf BrPic}(\cC)$, which include outer automorphisms of $\cC$ on top of the discrete gaugings ($\cC$-module categories) that produce $\cC$ as the dual category.\footnote{For example, the corresponding module category for trivial gauging (i.e. $A=\id$) is $\cC$ itself. However, as a $(\cC,\cC)$-bimodule category, $\cC$ may possess inequivalent bimodule actions of $\cC$ that come from composition with an outer-automorphism of $\cC$ from one side.}
It is obvious from the groupoid structure that all automorphism groups are isomorphic ${\rm \bf BrPic}(\cC)\cong {\rm \bf BrPic}(\cC')$ for $\cC,\cC'\in {\rm \bf BrPic}$.
Coming back to the case at hand, the ${\rm \bf Rep}(H_8)$ admits 6 inequivalent and indecomposable module categories, and 4 of them produce Morita autoequivalences of ${\rm \bf Rep}(H_8)$, while the remaining two produce Morita equivalence with the pointed category ${\rm \bf Vec}_{D_8}^\C$
(see Table~\ref{tab:repH8}). Consistency of sequential gauging further implies that the 4 autoequivalences generate a $\mZ_2\times \mZ_2$ subgroup of ${\rm \bf BrPic}({\rm \bf Rep}({H_8}))$, which together with the $\mZ_2$ outer-automorphism of ${\rm \bf Rep}(H_8)$ \eqref{outRepH8} gives,
\ie 
{\rm \bf BrPic}({\rm \bf Rep}({H_8}))=\mZ_2^3\,,
\label{brpicH8}
\fe
as is found in \cite{marshall2016brauerpicard}. The full groupoid ${\rm \bf BrPic}$ consists of $8\times 2^2=32$ invertible bimodule categories (Morita equivalence relations) whose composition laws are all commutative.

\subsection{Algebra objects in  ${\rm \bf Rep}(D_8)$}
We now follow the same strategy as in Section~\ref{sec:RepH8alg} to classify haploid algebra objects in ${\rm}(D_8)$. We find that, by going through the candidate algebras listed in Section~\ref{sec:NIMrepsTY} and solving the algebra conditions with the explicit ${\rm \bf Rep}(D_8)$ F-symbols in \eqref{TYFsymbol}, there are 12 different algebra objects which are listed in Table~\ref{tab:repD8}; all of them are Morita inequivalent except for the pair $A=\id,\cV\otimes \cV$ in the Morita trivial class.

The multiplication morphisms for $\mZ_2$ gauging $A=\id\oplus\eta$, $\id\oplus\eta'$, $\id\oplus\eta\eta'$ and $\mZ_2\times \mZ_2$ gauging $A=\id\oplus\eta\oplus \eta'\oplus \eta\eta'$ are the same as in \eqref{Z2mmap} (up to relabeling the legs of $m^k_{ij}$) and \eqref{Z2Z2mmap}, respectively, for ${\rm \bf Rep}(H_8)$ in our gauge fixing convention. 

For non-invertible gauging with algebra object $A=\id\oplus\eta\oplus \cV$ we find the following unique multiplication morphism,
\ie
m_{11}^1 = m_{1\eta}^{\eta} = m_{\eta 1}^{\eta} = m_{\eta \eta}^1 = m_{1\cV}^{\cV} = m_{\cV 1}^{\cV} = m_{\cV\cV}^1 = m_{\eta\cV}^{\cV} = m_{\cV\eta}^{\cV} = m_{\cV\cV}^{\eta} = \frac{1}{2} \,,
\label{D8dim4algebra}
\fe
and similarly for $\id\oplus\eta'\oplus \cV$ and $\id\oplus\eta\eta'\oplus \cV$ related by the $S_3$ triality automorphism in \eqref{outfusionring}. 

Finally for gauging the maximal algebra object $A=\id\oplus\eta\oplus\eta'\oplus\eta\eta'\oplus 2\cV$ the multiplication morphism is
\ie
\label{D8fibfunctors}
m_{g,h}^{gh} = \frac{1}{2\sqrt{2}} \begin{pmatrix}
    1 & 1 & 1 & 1 \\
     1 & 1 & -i & i \\
      1 & i & 1 & -i \\
       1 & -i & i & 1
\end{pmatrix} ,\, 
m_{g\cV}^{\cV} = (m_{\cV g}^{\cV})^t = m_{\cV\cV}^{g} = \frac{1}{2\sqrt{2}}(1_2, \sigma_1, \sigma_2, \sigma_3) \,,
\fe
and the other two solutions obtained by cyclically permuting the Pauli matrices $\sigma_i$ with $i=1,2,3$ in the last equality. These three solutions are again related by the $S_3$ automorphism \eqref{outfusionring}.

\begin{table}[!htb]
    \centering
    \begin{tabular}{|c|c|c|c|} \hline
      $L\subset D_8$ & Algebra object $A$   & Module Category $\cC_A$ & Dual Fusion Category ${}_A \cC_A$ \\\hline  \hline
1& $(\id\oplus\eta\oplus\eta'\oplus\eta\eta'\oplus 2\cV)_1$  & ${\rm NR}_6$: $\{A\}_1$ & ${\rm \bf Vec}_{D_8}$
     \\\hline 
   $\la  r^2\ra$ &  $(\id\oplus\eta \oplus\eta'\oplus \eta\eta')_\star$  
     & ${\rm NR}_5$: $\{A,2\cV\} $ &  ${\rm \bf Vec}_{\mZ_2^3}^\A$ 
         \\\hline 
    $\la  s\ra$  & $\id\oplus\eta\eta' \oplus\cV$  & ${\rm NR}_4$: $\{A,\eta\oplus   \eta'\oplus \cV\}$ & ${\rm \bf Rep}(D_8)$
           \\\hline 
    $\la  sr\ra$    & $\id\oplus\eta' \oplus\cV$  & ${\rm NR}_4$: $\{A,\eta\oplus \eta \eta'\oplus \cV\}$ & ${\rm \bf Rep}(D_8)$
      \\\hline 
     $\la  s,r^2\ra$ & $\id\oplus\eta\eta'$     &  ${\rm NR}_3$: $\{A,\eta\oplus  \eta',\cV_1,\cV_2\}$ &   ${\rm \bf Vec}_{D_8}$ 
        \\\hline 
     $\la  s,r^2\ra_\star$
     &  $(\id\oplus\eta\oplus\eta'\oplus\eta\eta'\oplus 2\cV)_2$  & ${\rm NR}_6$:  $\{A\}_2$ & ${\rm \bf Vec}_{D_8}$
     \\\hline  
     $\la  sr,r^2\ra$ & $\id\oplus\eta'$    & ${\rm NR}_3$: $\{A,\eta\oplus \eta \eta',\cV_1,\cV_2\}$  &  
${\rm \bf Vec}_{D_8}$       \\\hline 
     $\la  sr,r^2\ra_\star$
    &      $(\id\oplus\eta\oplus\eta'\oplus\eta\eta'\oplus 2\cV)_3$  & ${\rm NR}_6$: $\{A\}_3$ & ${\rm \bf Vec}_{D_8}$
     \\\hline 
       $\la  r\ra$   &    ${\id \oplus\eta }$  
    & ${\rm NR}_3$:   $\{A,\eta'\oplus \eta\eta',\cV_1,\cV_2\}$  &  ${\rm \bf Vec}_
    {D_8}$ 
    \\\hline 
     $D_8$  & $\id,\id\oplus\eta\oplus \eta'\oplus \eta\eta'$  & ${\rm NR}_1$: ${\rm \bf Rep}(D_8)$ & ${\rm \bf Rep}(D_8)$ \\\hline 
     ${D_8}_\star$
  &    ${\id\oplus   \eta\oplus \cV}$  
    &  ${\rm NR}_4$: $\{A,\eta'\oplus \eta\eta'\oplus \cV\}$ &   ${\rm \bf Rep}(D_8)$\\ \hline
    \end{tabular}
    \caption{Algebra objects $A$ in ${\rm \bf Rep}(D_8)$ grouped according to the Morita equivalence classes together with the corresponding module categories $\cC_A$ (simple objects therein and the corresponding NIM-rep from Section~\ref{sec:NIMrepsTY}) and dual fusion categories ${}_A\cC_A$. The subscripts on the algebra object $A$ label different multiplication morphisms. The first column labels the module categories using group-theoretical data in $D_8$ according to \eqref{gtmodulecat} for $(H,\psi)=(D_8,1)$.}
    \label{tab:repD8}
\end{table}

\subsection{Module Categories and Orbifold Groupoid for ${\rm \bf Rep}(D_8)$}
From the complete list of haploid algebras $(A,m)$ in ${\rm \bf Rep}(D_8)$ found in the last subsection, we can immediately read off the corresponding NIM-irreps from Section~\ref{sec:NIMrepsTY} and they are listed in Table~\ref{tab:repD8}.\footnote{We thank Yanming Su for pointing out a typo in the table. } Below, we will explain how we obtain the rest of the gauging data summarized in Table~\ref{tab:repD8} as well as the generalized orbifold groupoid for ${\rm \bf Rep}(D_8)$. 
We will be brief about similar steps that we have gone over in more detail for ${\rm \bf Rep}(H_8)$ in Section~\ref{sec:H8modulecat} and focus on the new features for ${\rm \bf Rep}(D_8)$. We first note that the F-symbols for ${\rm \bf Rep}(D_8)$ from \eqref{TYFsymbol} respects the $S_3$ automorphism of the fusion ring \cite{tambara2000representations},
\ie 
{\rm Out}({\rm \bf Rep}(D_8))=S_3\,,
\label{outRepD8}
\fe
which will play an important role in the structure of the relevant (bi)module categories.

\paragraph{Module categories for ${\rm \bf Rep}(D_8)$}
There are 11 indecomposable module categories for ${\rm \bf Rep}(D_8)$ corresponding to the 12 haploid algebra objects described in the last subsection, two of which are in the Morita trivial class, given by $A=\id,\cV\otimes \cV$, and produce the universal regular module category with NIM-irrep \eqref{NR1}. For the other module categories we list the simple objects as $A$-modules (using induced modules) in Table~\ref{tab:repD8} as we have done in Section~\ref{sec:H8modulecat}. One novelty here compared to the previous case is that, there are three inequivalent algebra structures on the maximal algebra object (see \eqref{amaxdef}) $A=\id\oplus\eta\oplus\eta'\oplus\eta\eta'\oplus 2\cV$ given by multiplication morphisms $m$ in \eqref{D8fibfunctors}, which correspond to the three inequivalent fiber functors of ${\rm \bf Rep}(D_8)$ permuted into each other by the $S_3$ triality \cite{tambara2000representations}. They produce three inequivalent rank 1 module categories with the same NIM-irrep \eqref{NR6} but differ in the module structure on ${\rm \bf Vec}$. Similarly the rank 2 module categories also come in a family of 3 related by $S_3$
with the same NIM-irrep \eqref{NR5}. On the other hand, the rank 4 module categories, which also come in a family of 3, are distinguished by their NIM-irreps \eqref{NR4} on which the $S_3$ triality acts nontrivially. Note that the other rank 4 NIM-irrep \eqref{NR2} does not give rise to a ${\rm \bf Rep}(D_8)$-module categories.

Since ${\rm \bf Rep}(D_8)$ is group-theoretical, we can compare our classification with existing results from group-theoretical methods \cite{ostrik2001module,10.1155/S1073792803205079,Natale_2017,etingof2019tensor} (see also related discussions in \cite{Thorngren:2019iar} based on the phases of the $D_8$ gauge theory). After realizing ${\rm \bf Rep}(D_8)$ as the dual category of ${\rm \bf Vec}_{D_8}$ under $D_8$ gauging as in \eqref{H8D8grouptheoretical}, we can identify module categories of ${\rm \bf Rep}(D_8)$ with those of ${\rm \bf Vec}_{D_8}$ by sequential gauging as described in Section~\ref{sec:seqgauginggrouptheoretical}. These module categories are labeled as $\cM^{D_8,1}(L,\xi)$ as in \eqref{gtmodulecat} for subgroups $L\subset D_8$ (up to conjugacy) and 2-cocycles $\xi\in H^2(L,\mZ_2)$,
\ie 
L_\xi=1\,,~\la s \ra\,,~\la sr \ra\,,~\la r^2\ra\,,~\la r \ra\,,~\la s,r^2 \ra_\xi\,,~\la sr,r^2 \ra_\xi\,,~{D_8}_\xi\,,
\label{Llistfull}
\fe
where $\xi$ is $\mZ_2$ valued (from $H^2(\mZ_2^2,U(1))=H^2(D_8,U(1))=\mZ_2$). Physically they correspond to mixed Dirichlet-Neumann boundary conditions for the $D_8$ gauge fields. Moreover the module category $\cM^{D_8,1}(L,\xi)$ is the category of projective representations of $L$ where the projective phase determined by $\xi$ and the corresponding NIM-irreps follow immediately from the fusion rules for such representations.
We thus establish the correspondence between group-theoretical data $L_\xi$ in \eqref{Llistfull} with the module categories in Table~\ref{tab:repD8}.\footnote{This involves choosing a specific $S_3$ triality frame.}

\paragraph{Dual fusion categories for ${\rm \bf Rep}(D_8)$}
The dual categories for generalized discrete gauging of ${\rm \bf Rep}(D_8)$ can also be determined from sequential gauging starting from the obvious relation with ${\rm \bf Vec}_{D_8}$ by gauging the entire $D_8$,
\ie 
{\rm \bf Rep}(D_8) = (({\rm \bf Vec}_{D_8})_{\cM(D_8,1)}^*)^{\rm op}\,,
\fe 
and the reverse gauging that returns ${\rm \bf Vec}_{D_8}$ by gauging the dual maximal algebra $A=(\id\oplus\eta\oplus\eta'\oplus\eta\eta'\oplus 2\cV)_1$ in Table~\ref{tab:repD8}, 
\ie 
{\rm \bf Vec}_{D_8} = (({\rm \bf Rep}({D_8})_{\cM^{D_8,1}(1,1)}^*)^{\rm op}\,.
\fe 
We can then go down the list of $L$ in Table~\ref{tab:repD8} to determine the dual fusion category ${}_A{\rm \bf Rep}(D_8)_A$ for each algebra $A$ and the corresponding $L$, using well-known results in gauging non-anomalous abelian subgroups of finite groups \cite{Bhardwaj:2017xup,Tachikawa:2017gyf,Chang:2018iay,Thorngren:2021yso}. When $L=D_8$ (with or without discrete torsion), we get 
back ${\rm \bf Rep}(D_8)$ as is evident from the above discussion. All other subgroups of $D_8$ are abelian.
When $L$ is a non-normal subgroup (i.e. $L=\la s\ra\,,\la sr\ra$), this discrete gauging again creates the non-invertible symmetry category ${\rm \bf Rep}(D_8)$, where the non-invertible duality TDL comes from the direct sum of TDLs in the nontrivial $L$-orbit. When $L$ is an abelian normal subgroup, then the dual symmetry is again a group of order 8 containing normal subgroup $\hat L\cong L$. The full group structure depends on whether $L$ gives a nontrivial extension of the quotient $D_8/L$ by $L$.\footnote{A group extension given by the short exact sequence
\ie 
1 \to N \to G\to Q \to 1\,,
\fe
is nontrivial if the extension class $H^2(Q,N)$ is taken to be nontrivial. Equivalently, a trivial group extension corresponds to a split exact sequence which means $G=N\rtimes Q$ is a semi-direct product (which includes the direct product $G=N\times Q$ as a special case).
}
Here for $D_8$ the only nontrivial extension arises for the exact sequence \eqref{D8extension}, in which case we obtain the well-known result that the dual symmetry is ${\rm \bf Vec}_{\mZ_2^3}^\A$ where the nontrivial extension leads to an anomaly characterized by the following 3-cocycle in $H^3(\mZ_2^3,U(1))=\mZ_2^7$  \cite{deWildPropitius:1995cf},\footnote{In the original context \cite{deWildPropitius:1995cf}, this explains the equivalence between the twisted $2+1d$ abelian Dijkgraaf-Witten gauge theory ${\rm DW}[\mZ_2^3]_\A$ (where $\A$ specifies the TQFT action and the untwisted non-abelian gauging theory ${\rm DW}[D_8]$.}
\ie 
\A(g_1,g_2,g_3)=e^{\pi i g_1g_2g_3}\,,\quad g_i\in \mZ_2\,.
\label{3cocycleZ23}
\fe
In all other cases where $L$ is a proper normal subgroup, the dual symmetry is clearly just ${\rm \bf Vec}_{D_8}$ itself. We have thus finished explaining the gauging data in Table~\ref{tab:repD8}.

\paragraph{Generalized orbifold groupoid for ${\rm \bf Rep}(D_8)$}
Compared to the case of ${\rm \bf Rep}(H_8)$, the ${\rm \bf Rep}(D_8)$ fusion category admits a richer gauging structure which is reflected in its generalized orbifold groupoid ${\rm \bf BrPic}$. In particular the 11 module categories in Table~\ref{tab:repD8} lead to 4 Morita auto-equivalences for ${\rm  Rep}(D_8)$, 6 Morita equivalences between ${\rm  Rep}(D_8)$ and ${\rm \bf Vec}_{D_8}$, and 1 Morita equivalence between ${\rm  Rep}(D_8)$ and ${\rm \bf Vec}_{\mZ_2^3}^\A$. These 4 Morita auto-equivalences again form a $\mZ_2\times \mZ_2$ group and after combining with the $S_3$ outer-automorphism \eqref{outRepD8} which permute the three $\mZ_2$ generators (see Table~\ref{tab:repD8}) they together produce 24 auto-equivalences (i.e. invertible $({\rm \bf Rep}(D_8),{\rm \bf Rep}(D_8))$-bimodule categories) that form the Brauer-Picard group
\ie 
{\rm \bf BrPic}({\rm \bf Rep}({D_8}))=S_4\,,
\fe
reproducing the result of 
\cite{NIKSHYCH2014191}.

\begin{figure}[!htb]
    \centering
   \includegraphics[scale=2.25]{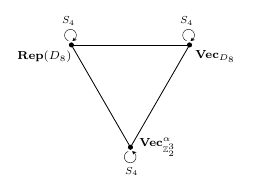}
    \caption{Generalized orbifold groupoid for ${\rm \bf Rep}(D_8)$.}
    \label{fig:D8groupoid}
\end{figure}

The full (connected) groupoid ${\rm \bf BrPic}$ is presented in Figure~\ref{fig:D8groupoid}. It consists of three nodes representing the Morita equivalent fusion categories ${\rm \bf Rep}(D_8)$, ${\rm \bf Vec}_{D_8}$, and ${\rm \bf Vec}_{\mZ_2^3}^\A$ connected by Morita equivalence relations (invertible bimodule categories). There are a total of $24\times 3^2=216$ invertible bimodule categories.

\section{Examples of Generalized Gauging in CFT}
\label{sec:CFTrealization}

In this section, we apply what we have learned in Section~\ref{sec:genprop} to concrete examples of generalized gauging in 2d CFTs. 
 
\subsection{Ising$^2$ CFT and Infinite Non-invertible Self-duality}
\label{sec:Isingsq}

The first example we study is one of the simplest nontrivial CFTs at $d=2$; it is the theory constructed from the tensor product of two copies of the Ising CFT, which we refer to as the Ising$^2$ CFT. As was analyzed in \cite{Thorngren:2019iar,Thorngren:2021yso}, despite its simplicity, this CFT hosts a rich zoo of non-invertible symmetries that provide an ideal playground to study generalized gauging of non-invertible symmetries. 

\subsubsection{Operator Content and Verlinde TDLs in ${\rm Ising}^2$ CFT}
We start by reviewing the operator content of the Ising$^2$ CFT and its generalized symmetries. The Ising$^2$ CFT has conformal central charge $c=1$ and is a rational CFT (RCFT) with respect to the chiral algebra $V_{{\rm Ising}^2}\equiv {\rm Vir}_{c={1\over 2}}\times {\rm Vir}_{c={1\over 2}}$. The primary operators with respect to $V_{{\rm Ising}^2}$ are given by tensor products of those in each Ising factor,
\ie 
\cH^{\rm primaries}_{S^1}=\{\id, \ep_1,\sigma_1\}\otimes \{\id, \ep_2,\sigma_2\}\,,
\fe
where $\ep_i$ is the $h=\bar h={1\over 2}$ energy operator and $\sigma_i$ is the $h=\bar h={1\over 16}$ spin operator. The full operator content (including the descendants) of the Ising$^2$ CFT is captured by its torus partition function
\ie 
Z({\rm Ising}^2)=
(|\chi_\id|^2 +|\chi_{\ep}|^2+|\chi_\sigma|^2)^2\,,
\fe
where $\chi_\id,\chi_\ep,\chi_\sigma$ are the usual chiral Ising characters.

As an RCFT, the obvious symmetries come from the Verlinde TDLs, 
which generate a subcategory of the full symmetry category $\cC_{{\rm Ising}^2}$,
\ie 
{\rm  \bf Ising}\boxtimes{\rm \bf Ising} \subset \cC_{{\rm Ising}^2}\,,
\label{Isingsqver}
\fe
from the symmetry category of each Ising factor
\ie 
\cC_{\rm Ising}={\rm  \bf Ising}=\{\id,\eta,\cN\}\,,\quad \cN^2=\id \oplus\eta\,,~\eta^2=\id\,,~\cN\eta=\eta\cN=\cN\,.
\label{Isingcat}
\fe
However, unlike a single Ising factor whose symmetry category is finite and fully determined, the Ising$^2$ CFT has many more symmetries due to the much richer operator content; in fact, $\cC_{{\rm Ising}^2}$ contains uncountably infinitely many simple TDLs \cite{Thorngren:2019iar,Chang:2020imq,Thorngren:2021yso}. 

\subsubsection{Infinite Generalized Symmetries on $c=1$ Orbifold
Branch}

The Ising$^2$ CFT comes with an exactly marginal parameter that generates a one-dimensional conformal manifold which corresponds to the orbifold branch of the $c=1$ CFT. Such CFTs are known to be described by a free compact boson $\vartheta\sim \vartheta+2\pi$ with the following action
\ie 
S={R^2\over 8\pi}\int  d^2 x \,\pa_\m \vartheta \pa^\m \vartheta\,,
\label{freeboson}
\fe
and identification (orbifold) by the $\mZ_2^C$ symmetry that sends $\vartheta$ to $-\vartheta$. The radius $R$ in \eqref{freeboson} parametrize the continuous orbifold branch of the conformal moduli space at $c=1$ \cite{Ginsparg:1987eb}. T-duality leads to an equivalence relation among the orbifold CFTs by
\ie 
R \leftrightarrow {2\over R}\,,\quad \vartheta \leftrightarrow \varphi\,,
\label{Tdual}
\fe 
where $\varphi$ denotes the dual compact boson with the same $2\pi$ periodicity; therefore, the inequivalent CFTs on the orbifold branch can be parametrized by $R\geq \sqrt{2}$ and the T-duality fixed point $R=\sqrt{2}$ represents the famous Berezinskii–Kosterlitz–Thouless (BKT) point where another continuous branch of the $c=1$ moduli space emerges that describes the compact boson \eqref{freeboson} without orbifolding (also known as the circular branch which is again parametrized by $R\geq \sqrt{2}$ from \eqref{freeboson}). 
 
The CFTs on the $c=1$ circular branch share an obvious invertible global symmetry\footnote{This is believed to be the full symmetry at a generic point on the circular branch of the $c=1$ moduli space.}
\ie 
G_{\rm circ}=(U(1)_\vartheta\times U(1)_\varphi) \rtimes \mZ_2^C\,,
\label{Gcirc}
\fe
where the two $U(1)$ factors act as periodic shift symmetries in the (dual) boson fields $\vartheta$ and $\varphi$, while the charge conjugation $C$ (which generates $\mZ_2^C$)  acts by reflection on $\vartheta$ and $\varphi$. Gauging $\mZ_2^C$ sends the circular branch CFTs to the orbifold branch at the same $R$. Since $\mZ_2^C$ is a non-normal subgroup of $G_{\rm circ}$, this creates non-invertible symmetries in the gauged theory by summing over $\mZ_2^C$ orbits of invertible TDLs in $G_{\rm circ}$. The resulting non-invertible TDLs have quantum dimension 2 and come in a two-parameter family first studied in \cite{Chang:2020imq},\footnote{All such TDLs $\cL_{\theta,\phi}$ are clearly self-dual and have the trivial Frobenius-Schur indicator (which can be understood either from the $\mZ_2^C$ gauging on the circular branch or by continuity from the trivial Frobenius-Schur indicators at $\theta,\phi=0,\pi$).\label{ft:FScont}}
\ie 
\cC_{\rm cont}=&\{\cL_{\theta,\phi} \,|\,\cL_{\theta,\phi}\otimes \cL_{\theta',\phi'}=\cL_{\theta+\theta',\phi+\phi'}\oplus \cL_{\theta-\theta',\phi-\phi'}\}\,,
\label{contTDL}
\fe
where the parameters $\theta,\phi$ are $2\pi$-periodic and are identified by the $\mZ_2^C$ reflection via $(\theta,\phi)\sim (-\theta,-\phi)$. 
This continous family of TDLs are simple except at $\theta,\phi=0,\pi$,
\ie 
\cL_{0,0}=\id \oplus r^2\,,\quad \cL_{0,\pi}=s \oplus sr^2
\,,\quad \cL_{\pi,0}=sr \oplus sr^3\,,\quad \cL_{\pi,\pi}=r \oplus r^3\,,
\label{thetaphisplitting}
\fe
where $s,r$ generate the $D_8$ invertible symmetry (see \eqref{D8}) on the orbifold branch that comes from the nontrivial extension of the commutant $\mZ_2^\vartheta\times \mZ_2^\varphi\subset G_{\rm circ}$ of $\mZ_2^C$  by the dual (magnetic) $\hat{\mZ}_2$ symmetry as a consequence of the nontrivial mixed anomaly (as in \eqref{3cocycleZ23}) for $\mZ_2^\vartheta\times \mZ_2^\varphi \times \mZ_2^C$ \cite{Thorngren:2021yso}. 
Consequently the dual symmetry to $\cC_{\rm circ}$ given by the symmetry group \eqref{Gcirc} on the orbifold branch is,
\ie 
\cC_{\rm orb}={\cC_{\rm cont}  \boxtimes {\rm \bf Vec}_{D_8} /\sim}\,,
\label{Corb}
\fe
where the quotient implements the identifications in \eqref{thetaphisplitting}.\footnote{There are further enhancements at special points on the orbifold branch \cite{Thorngren:2019iar,Thorngren:2021yso}.}  We emphasize that, just as \eqref{Gcirc} is the symmetry shared by all points on the circular branch, \eqref{Corb} is preserved on the entire orbifold branch (and plausibly the full symmetry category at generic $R$).
The appearance of two continuous families of non-invertible TDLs is also expected intuitively since discrete (generalized) gauging preserves the total quantum dimension of the symmetric category (in the finite case), and the case here can be thought of as coming from a certain direct limit of fusion categories (and their duals) \cite{Thorngren:2021yso}.

The infinite non-invertible symmetry $\cC_{\rm orb}$ \eqref{Corb} contains many interesting fusion categories. In particular the ${\rm \bf Rep}(H_8)$ and ${\rm \bf Rep}(D_8)$ symmetries we have studied in detail in Section~\ref{sec:gengaugingRep} are fusion subcategories of $\cC_{\rm orb}$. We can identify the non-invertible TDLs thereof using TDLs in $\cC_{\rm orb}$,
\ie 
\cV_{{\rm \bf Rep}(D_8)}=\cL_{{\pi \over 2},0}\,,\quad \cV_{{\rm \bf Rep}(H_8)}=s\cL_{{\pi \over 2},0}=\cL_{{\pi \over 2},\pi}\,,
\quad \eta=sr\,,\quad \eta'=rs\,,\quad \eta\eta'=r^2\,,
\label{identifyRepfromcont}
\fe
following the same convention as in \cite{Thorngren:2021yso}. There is another copy of ${\rm \bf Rep}(H_8)$ and ${\rm \bf Rep}(D_8)$ in $\cC_{\rm orb}$ related to the above by T-duality \eqref{Tdual}. By taking $\cL_{{2\pi \over k},0}$ (or $\cL_{0,{2\pi \over k}}$) for general positive integer $k\geq 4$, we can generate copies of ${\rm \bf Rep}(D_{2k})$ in $\cC_{\rm orb}$. More generally, by considering the subcategories generated by $\cL_{{2\pi p\over k},{2\pi q\over  k}}$ for $p,q\in \mZ_k$ and coprime with $k$, we obtain higher rank analogs of ${\rm \bf Rep}(H_8)$ and generalizations.
Finally we also note that the Tambara-Yamagami symmetry ${\rm \bf TY}(\mZ_4,\chi_+,\ep=+)$ (and another one with the conjugate bicharacter $\chi_-$) on the orbifold branch  \cite{Thorngren:2021yso} that explains the self-duality under gauging the $\mZ_4=\la r\ra$ subgroup of $D_8$ \cite{Yang:1987wk} is also contained in $\cC_{\rm orb}$ and generated by $\cL_{{\pi\over 2},\pm {\pi\over 2}}$.

\subsubsection{General Fusion Category Symmetries in Ising$^2$ CFT}

Coming back to the Ising$^2$ CFT that sits at the bosonization/fermionization radius $R=2$ on the $c=1$ orbifold branch, we note the symmetry is further enhanced beyond \eqref{Corb} by \eqref{Isingsqver} \cite{Thorngren:2021yso}, and given the quotient of a bi-crossed product (or an exact
factorization) category \cite{gelaki2017exact,natale2018classification},
\ie 
\cC_{{\rm Ising}^2} \supset ({\rm  \bf Ising}\boxtimes{\rm \bf Ising}) \Join \cC_{\rm orb} /\sim \,,
\label{Isingsqsym}
\fe
where the ${\rm \bf Vec}_{D_8}$ subcategory of $\cC_{\rm orb}$ acts on the Ising factors by an $\mZ_2$ (exchange) automorphism and the identification is between the common ${\rm \bf Rep}(H_8)$ subcategory of ${\rm  \bf Ising}\boxtimes{\rm \bf Ising}$ and $\cC_{\rm orb}$ given in \eqref{identifyRepfromcont}. In particular we have the following relation between the duality TDLs $\cN,\cN'$ in the Ising categories \eqref{Isingcat} and those from the continuous families \eqref{contTDL} \cite{Chang:2020imq,Thorngren:2021yso},
\ie 
\cL_{{\pi \over 2},\pi} =  \cN \cN'=\cV_{{\rm \bf Rep}({H_8})}\,,
\label{idRepH8Isingsqcont}
\fe 
 We will also make use of the following fusion rules between the Ising duality TDLs and those in $\cC_{\rm orb}$,
 \ie 
s\cN=\cN's\,,\quad \cN \otimes \cL_{\theta,\phi}=\cL_{{\phi\over 2},2\theta} \otimes \cN\,,
\label{usefulfusion}
 \fe
where the second equality follows T-duality on the orbifold branch (which establishes the isomorphism between the Ising$^2$ and its $\mZ_2$ gauging by $\cN^2=\id\oplus \eta$).

\subsubsection{Generalized Gauging in Ising$^2$ CFT and Infinite Non-invertible Self-duality}

Let us now consider generalized gauging in the Ising$^2$ CFT using the symmetries identified in \eqref{Isingsqsym}. For brevity, we mostly focus on two subcategories, namely ${\rm \bf Rep}(H_8)$ and ${\rm \bf Rep}(D_8)$ for which we have worked out their universal properties concerning generalized gauging in Section~\ref{sec:gengaugingRep}, and we will now see how they are realized in a nontrivial physical CFT.

We start with gauging the ${\rm \bf Rep}(H_8)$ symmetry of Ising$^2$ CFT. The possible distinct gaugings are classified in Table~\ref{tab:repH8}. We note immediately that 6 of the 8 algebra objects $(A,m)$ in ${\rm \bf Rep}(H_8)$ become Morita trivial (with respect to a larger but finite subcategory) in the symmetry category $\cC_{{\rm Ising}^2}$ of the Ising$^2$ CFT, because of the existence of simple TDLs in $\cC_{{\rm Ising}^2}$ that ``square'' to $A$,
\ie 
&\cN^2=\id\oplus \eta\,,\quad \cN'^2=\id\oplus \eta'\,,\quad \cV_{{\rm \bf Rep}({H_8})}^2=\id\oplus \eta\oplus \eta'\oplus \eta\eta'\,, 
\\
&\cL_{{\pi \over 4},{\pi \over 2}}^2= \id \oplus \eta\eta' \oplus \cV_{{\rm \bf Rep}({H_8})}\,,\quad (\cN\cL_{{\pi \over 4},{\pi \over 2}})^2= \id \oplus \eta\oplus \eta'  \oplus\eta\eta' \oplus 2\cV_{{\rm \bf Rep}({H_8})}\,,
\label{moritatrivialIsingsq}
\fe
where we have used the identification \eqref{idRepH8Isingsqcont} and \eqref{usefulfusion}. The only Morita nontrivial algebras are two invertible gaugings,
\ie 
A=\id \oplus \eta \eta'\,,\quad (\id \oplus \eta\oplus \eta'\oplus \eta \eta')_\star\,,
\label{isingsqexceptA}
\fe
which are Morita equivalent because (see around \eqref{moritaequivalent})
\ie 
\cN \otimes (\id \oplus \eta\eta')\otimes \cN =\id \oplus \eta \oplus \eta'\oplus \eta\eta'\,,
\fe
with an induced multiplication map from the RHS that is different from $\cV_{{\rm \bf Rep}({H_8})}^2$ in \eqref{moritatrivialIsingsq}, which is Morita trivial and thus must coincide with that of $A=(\id \oplus \eta\oplus \eta'\oplus \eta \eta')_\star$ in Table~\ref{tab:repH8}. 

 Therefore, all except \eqref{isingsqexceptA} gauging of ${\rm \bf Rep}(H_8)$ in the Ising$^2$ CFT necessarily produce self-duality.\footnote{In particular, \ie \cN\cL_{{\pi \over 4},\mp{\pi \over 2}}\cong  \cN\cL_{{3\pi \over 4},\pm {\pi \over 2}}\,,
 \label{RepH8ext}
 \fe are the two self-duality TDLs $\cD,\cD'$ identified in the recent work \cite{Choi:2023vgk} for gauging the unique maximal algebra in a specific ${\rm \bf Rep}(H_8)$ symmetry of the ${\rm Ising}^2$ CFT \cite{Thorngren:2021yso}. They are related by fusing with the invertible generator $s\in D_8$ that swaps the two Ising factors and they generate two inequivalent fusion categories as $\mZ_2$ extensions of the ${\rm \bf Rep}(H_8)$. Such extensions of a fusion category $\cC$ by $G$ are classified 
using the Brauer-Picard groupoid of $\cC$ \cite{etingof2009fusion}. More explicitly, the inequivalent $G$-extensions correspond to
 homotopy classes of maps between classifying spaces $BG \to B\underline{\underline{\rm \bf BrPic}}(\cC)$ (which consists of a group homomorphism $\rho: G \to  {\rm \bf BrPic}(\cC)$ together with a suitable 2-cocycle and a 3-cocycle for $G$ that satisfy vanishing obstruction conditions) \cite{etingof2009fusion}. Here for $G=\mZ_2$ and $\cC={\rm \bf Rep}(H_8)$, with the Brauer-Picard groupoid given in Section~\ref{sec:H8modulecat} (see around \eqref{brpicH8}), there are 8 inequivalent fusion categories from the $\mZ_2$-extension of $\cC={\rm \bf Rep}(H_8)$ parametrized by four distinct choices of $\rho:\mZ_2\to \mZ_2^3$ and the Frobenius-Schur indicator $\ep_\cD$ of $\cD$ \cite{Choi:2023vgk}. In the Ising$^2$ CFT, $\ep_\cD=1$ (see Footnote~\ref{ft:FScont}) and two of the remaining four fusion categories (generated by $\cD$ and $\cD'$ in \cite{Choi:2023vgk}) are in one-to-one correspondence with those in \eqref{RepH8ext}. We thank the authors of \cite{Choi:2023vgk} for a correspondence that led to a  corrected statement above.} It is straightforward to check this explicitly at the level of torus partition functions using the explicit form of the multiplication morphism given in Section~\ref{sec:gengaugingRep}.

One can carry out the parallel analysis for gauging the ${\rm \bf Rep}(D_8)$ symmetry of the Ising$^2$ CFT, whose algebra objects are listed in Table~\ref{tab:repD8}. The 11 Morita equivalence classes of algebra objects for ${\rm \bf Rep}(D_8)$ descend to 4 Morita equivalence classes with respect to a larger symmetry subcategory. Among them, the Morita trivial class  contains, in addition to those in \eqref{moritatrivialIsingsq} and the trivial class in Table~\ref{tab:repD8}, the following algebras 
\ie 
A= \id \oplus  \eta\eta'\oplus  \cV_{{\rm \bf Rep}({D_8})}\,,\quad (\id \oplus \eta\oplus \eta'\oplus\eta\eta'\oplus 2\cV_{{\rm \bf Rep}({D_8})})_{1,2}\,,\quad 
\fe
that are nontrivial in the ${\rm \bf Rep}(H_8)$ subcategory but trivializes in a larger subcategory of $\cC_{{\rm Ising}^2}$. The corresponding self-duality TDLs are $\cL_{{\pi\over 4},0}\,,\cL_{{\pi \over 4},0} \cN$, and $\cL_{{\pi \over 4},0} \cN'$, respectively,
\ie 
&\cL_{{\pi \over 4},0}^2= \id \oplus \eta\eta' \oplus \cV_{{\rm \bf Rep}({D_8})}\,,\quad  
\\
&  \cL_{{\pi \over 4},0} \cN \otimes  \cN \cL_{{\pi \over 4},0}  =
\cL_{{\pi \over 4},0} \cN' \otimes  \cN' \cL_{{\pi \over 4},0}= \id \oplus \eta \oplus \eta' \oplus \eta\eta' \oplus 2\cV_{{\rm \bf Rep}({D_8})}\,, 
\label{moritatrivialIsingsq2}
\fe
where we have used \eqref{usefulfusion} and\footnote{Note that it follows from \eqref{usefulfusion} that $\cL_{{\pi \over 4},0} \cN=\cN\cL_{0,{\pi\over 2}}$.}
\ie 
\cL_{0,{\pi\over 2}}^2=  \id \oplus s \oplus r^2 \oplus sr^2 \,.
\fe 
Note that the self-duality TDLs in \eqref{moritatrivialIsingsq2} are no longer self-dual unlike in \eqref{moritatrivialIsingsq}.

The first nontrivial Morita class in $\cC_{{\rm Ising}^2}$ from algebras of ${\rm \bf Rep}(D_8)$ coincides with \eqref{isingsqexceptA}.
The second nontrivial Morita class contains the following algebras
\ie 
A= \id \oplus  \eta \oplus  \cV_{{\rm \bf Rep}({D_8})}\,,\quad \id \oplus  \eta' \oplus  \cV_{{\rm \bf Rep}({D_8})}\,,
\label{moritanontrivialD82}
\fe
and the Morita equivalence relation follows from (as in \eqref{moritaequivalent})
\ie 
\cN\otimes  (\id \oplus s)\otimes \cN=\id \oplus  \eta \oplus  \cV_{{\rm \bf Rep}({D_8})}\,,\quad \cN'\otimes  (\id \oplus s)\otimes \cN'=\id \oplus  \eta' \oplus  \cV_{{\rm \bf Rep}({D_8})}\,.
\fe
The final nontrivial Morita class contains a single algebra from Table~\ref{tab:repD8},
\ie 
A=(\id \oplus \eta\oplus \eta'\oplus\eta\eta'\oplus 2\cV_{{\rm \bf Rep}({D_8})})_{3}\,,
\label{moritanontrivialD83}
\fe
which is Morita equivalent to gauging $A=\id \oplus s \oplus r^2 \oplus sr^2$ without discrete torsion.

It is clear that the shear amount of simple TDLs in $\cC_{\rm orb}$ creates many more Morita trivial algebras (with respect to a fusion subcategory in $\cC_{{\rm Ising}^2}$), such as the following
\ie 
A=\cL_{\theta,\phi}^2\,,~(\cN\cL_{\theta,\phi})\otimes (\cL_{\theta,\phi}\cN)\,,~(\cN'\cL_{\theta,\phi})\otimes (\cL_{\theta,\phi}\cN')\,,
\label{infiniteselfdualgauging}
\fe
for general $\theta,\phi$. They lead to infinitely many non-invertible self-dualities under generalized gauging in the ${\rm Ising}^2$ CFT (and also on the entire orbifold branch for the first type of algebras in \eqref{infiniteselfdualgauging}). Curiously it also includes a self-duality for invertible but non-abelian gauging,
\ie 
\cN\cL_{{\pi \over 4},0} \otimes \cL_{{\pi \over 4},0} \cN= \bigoplus_{g\in D_8} g\,,
\fe
which explains the self-duality of the ${\rm Ising}^2$ CFT under gauging the entire $D_8$ invertible symmetry \cite{Thorngren:2021yso}.

To complete this analysis, let us discuss what happens when we gauge the Morita nontrivial algebras in $\cC_{{\rm Ising}^2}$.
Under gauging either algebras in \eqref{isingsqexceptA} (which are Morita equivalent), the Ising$^2$ CFT turns into the compact boson CFT on the circular branch at the same bosonization/fermionization radius $R=2$, which is identified with the Dirac fermion CFT upon bosonization (GSO projection).\footnote{The bosonic Dirac CFT is also known as the $U(1)_4$ CFT since it is an RCFT with respect to the $U(1)_4$ chiral algebra.} The dual symmetry for gauging $A=\id \oplus \eta\eta'$ in ${\rm \bf Rep}(D_8)$ is ${\rm \bf Vec}_{D_8}$ from Table~\ref{tab:repD8}, which is an obvious non-anomalous $D_8$ subgroup of the invertible symmetry \eqref{Gcirc} on the circular branch generated by the $\mZ_4$ subgroup of $U(1)_\vartheta$ together with $\mZ_2^C$. On the other hand, gauging $A=(\id \oplus \eta \oplus \eta' \oplus\eta\eta')_\star$ in Table~\ref{tab:repD8} generates the anomalous ${\rm \bf Vec}_{\mZ_2^3}^\A$ symmetry that is identified with the $\mZ_2^\vartheta\times \mZ_2^\varphi\times \mZ_2^C$ subgroup of \eqref{Gcirc}, which has a mixed anomaly that involves all three $\mZ_2$ factors \cite{Thorngren:2021yso}. Furthermore, gauging the same algebras in ${\rm \bf Rep}(H_8)$ produces an anomalous dual symmetry ${\rm \bf Vec}_{D_8}^\C$ (see Table~\ref{tab:repH8}) that is identified with another $D_8$ subgroup of \eqref{Gcirc} generated by $\mZ_2^C$ and a $\mZ_4$ translation $\vartheta$ mixed with a $\mZ_2$ translation in $\varphi$.

Finally, under gauging the extra Morita nontrivial class of algebras from ${\rm \bf Rep}(D_8)$ in \eqref{moritanontrivialD82}, the Ising$^2$ CFT is mapped to another point on the orbifold branch at $R=4$, which corresponds to the $U(1)_8/\mZ_2$ CFT, using the equivalence to the invertible gauging of $\mZ_2^s$ \cite{Thorngren:2021yso}. The dual category ${\rm \bf Rep}(D_8)$ is the familiar symmetry on the entire orbifold branch \cite{Thorngren:2021yso}. 
When gauging the last Morita nontrivial class in \eqref{moritanontrivialD83}, we obtain instead the $U(1)_8$ CFT on the circular branch at $R=4$, where the dual category ${\rm \bf Vec}_{D_8}$ is the same non-anomalous $D_8$ symmetry that come from gauging $A=\id \oplus \eta\eta'$ above (but at a different radius). It is interesting to see how the above discussions are consistent with sequential gauging as there are many possible gauging sequences, especially in the ${\rm \bf Rep}(D_8)$ case, thanks to the rich groupoid structure.

\subsection{Generalized Gauging and Orbifold Groupoid in Irrational CFT}

Generalized gauging in rational CFTs can be understood largely in an algebraic manner in terms of the underlying chiral algebra $\cA$, equivalently the vertex operator algebra (VOA), its representation category ${\rm \bf Rep}(\cA)$, which defines a modular tensor category (MTC), and condensations of anyons in the MTC \cite{Fuchs:2002cm}. However, the situation is much more complicated for irrational CFTs, which do not have a finite number of conformal blocks with respect to any chiral algebra. It is expected that generic CFTs are irrational; this is certainly the case on the $c=1$ conformal moduli space, where every point on either the circular or orbifold branches such that $R^2\notin {\mathbb Q}$ is an irrational CFT. 
Nonetheless, the properties of generalized gauging that we have discussed at length in Section~\ref{sec:genprop} continue to apply, as they only rely on the basic axioms of a compact unitary local CFT.
Here we use the specific example of the $c=1$ orbifold branch to illustrate this point.

\begin{figure}[!htb]
    \centering
   \includegraphics[scale=1]{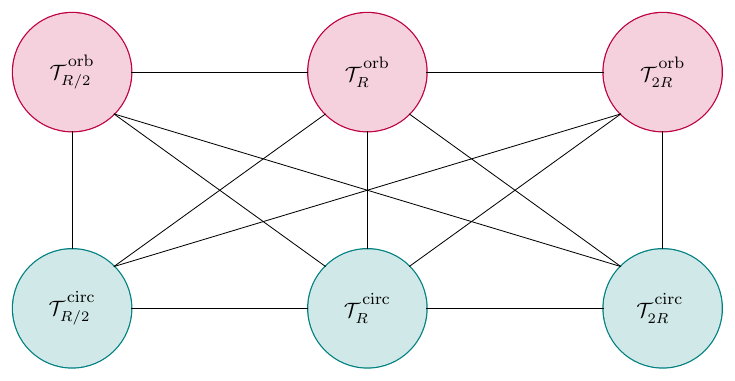}
    \caption{Generalized orbifold groupoid for the $c=1$ orbifold branch.}
    \label{fig:Orbgroupid}
\end{figure}

We first note that from the analysis in the last subsection, all algebra objects in ${\rm \bf Rep}(H_8)$ and ${\rm \bf Rep}(D_8)$ as listed in Table~\ref{tab:repH8} and Table~\ref{tab:repD8}, respectively, are Morita equivalent to gauging invertible finite subcategories in the symmetry category $\cC_{\rm orb}$ \eqref{Corb} that is preserved along the entire orbifold branch. For example, the unique maximal algebra of ${\rm \bf Rep}(H_8)$ satisfies,
\ie 
\id \oplus \eta\oplus \eta'\oplus\eta\eta'\oplus 2\cV_{{\rm \bf Rep}({H_8})}=\cL_{{\pi\over 4},{\pi \over 2}} \otimes (\id \oplus \eta )\otimes \cL_{{\pi\over 4},{\pi \over 2}} \,,
\fe 
and thus is equivalent to gauging the $\mZ_2 \subset D_8$ subgroup generated by $\eta=sr$. 
The rest of the gaugings in Table~\ref{tab:repH8} follow from sequential gauging with invertible symmetries. Similarly, the top maximal algebra of ${\rm \bf Rep}(D_8)$ in Table~\ref{tab:repD8} satisfies,\footnote{We have chosen a particular $S_3$ triality frame for this to hold as algebra objects, where the RHS carries the canonical algebra structure for objects of the form $\overline \cL \otimes A\otimes \cL$.}
\ie 
\id \oplus \eta\oplus \eta'\oplus\eta\eta'\oplus 2\cV_{{\rm \bf Rep}({D_8})}=\cL_{{\pi\over 4},0} \otimes (\id \oplus \eta )\otimes \cL_{{\pi\over 4},0} \,,
\fe 
and thus is again equivalent to gauging the same $\mZ_2$ mentioned above. Gauging any other algebras in the rest of Table~\ref{tab:repD8} follow from sequential gauging with invertible symmetries in $D_8$.  

\begin{figure}[!htb]
    \centering
  \includegraphics[scale=1]{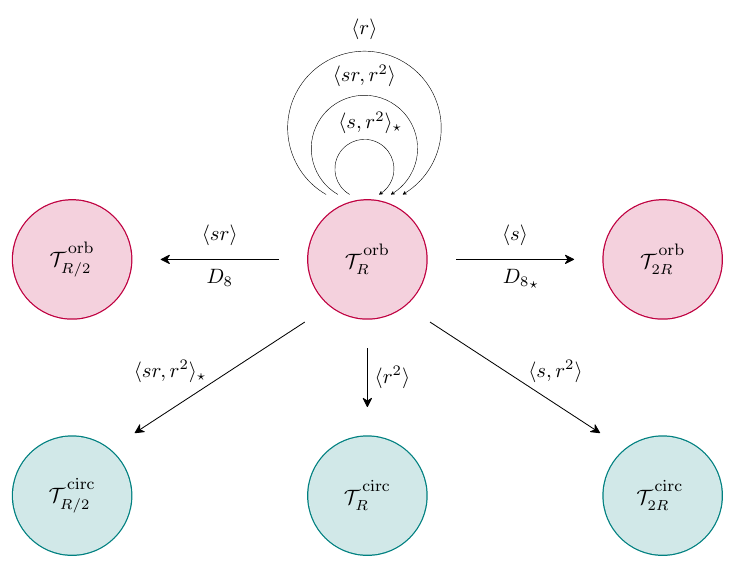}
    \caption{Explicit invertible bimodules from invertible gauging. The $\star$ subscript indicates nontrivial discrete torsion.}
    \label{fig:Orbgroupid2}
\end{figure}

The invertible gaugings on the $c=1$ orbifold branch by subgroups of $D_8$ were studied extensively in \cite{Thorngren:2021yso}, which connected different points on the orbifold and the circular branch where $R$ differed by a factor of 2, as illustrated in Figure~\ref{fig:Orbgroupid2}. Combining this with the understanding that all generalized gaugings on the orbifold branch are Morita equivalent to invertible gauging, we have thus identified the generalized orbifold groupoid (see Figure~\ref{fig:Orbgroupid}) for these irrational CFTs (at least for the algebra objects from ${\rm \bf Rep}(H_8)$ and ${\rm \bf Rep}(D_8)$).

\subsection{Binary Algebras and Gauging in Wess-Zumino-Witten CFT}
\label{sec:binaryalgebraWZW}
\subsubsection{Binary Algebras}
\label{sec:binaryintro}

Perhaps the simplest type of nontrivial haploid algebras one can imagine is an object of the following form,
\ie 
A=\id \oplus \cL_i\,,
\label{binaryalgebra}
\fe
which involves a single nontrivial self-dual TDL $\cL_i\in \cC$. We refer to such an algebra object with a multiplication morphism that solves the full set of consistency conditions as a \textit{binary algebra}.

For a binary algebra object \eqref{binaryalgebra}, the quadratic equations from the separability and associativity conditions are much simplified, and one can explicitly obtain the necessary and sufficient conditions on the F-symbols of the underlying fusion $\cC$ for the algebra to exist.

For example, when $\cL_i$ does not admit a trivalent topological junction with itself (i.e $N_{ii}^i=0$), the only F-symbol involved in the algebra conditions is (in the usual gauge)
\ie
{F_{iii}^{i}}(\id,\id) = \frac{\epsilon_i}{\la \cL_i \ra} \,,
\fe
and the associativity condition requires $F_{iii}^{i}(\id,\id)=1$, so $A$ can only be an algebra object when it is invertible and its Frobenius-Schur (FS) indicator is unity, $\epsilon_i= 1$. Such an algebra object corresponds to gauging a $\mZ_2$ symmetry generated by $\cL_i$, and its FS indicator then measures its 't Hooft anomaly \cite{Chang:2018iay}. We have thus recovered the familiar obstruction to gauging a $\mZ_2$ symmetry.

Now let's consider the next simplest case where the trivalent junction vector space for $\cL_i$ has dimension $N_{ii}^i=1$. The associativity constraint can be written as follows, where we have fixed the gauge such that $m_{i\id }^i = m_{\id i}^i = m_{ii}^{\id} = \frac{1}{\la A \ra^{1/2}}$ for any $i\in \cC$,
\ie
1 = F_{iii}^i(\id,\id) + F_{iii}^i(\cL_i,\id) \la A \ra (m_{ii}^i)^2\,, \\
\la A \ra (m_{ii}^i)^2 = F_{iii}^i(\id,\cL_i) + F_{iii}^i(\cL_i,\cL_i) \la A \ra (m_{ii}^i)^2\,, \\
F_{iii}^i(\id, \cL_j) + F_{iii}^i(\cL_i,\cL_j) \la A \ra (m_{ii}^i)^2 = 0 \,, 
\label{binary_qua_eqn}
\fe
where $j \neq i, \id$. Taking the F-symbols to be the input, one can derive the multiplication morphism (or prove its non-existence) from these simple quadratic equations. On the other hand, assuming we already know the existence of a multiplication morphism for the binary algebra object \eqref{binaryalgebra}, they can be used to derive a subset of the F-symbols without solving the pentagon equations. 

One of the simplest binary algebras for non-invertible gauging occurs for the ${\rm \bf Fib}$ category with simple TDLs $\{\id, W\}$. The unique nontrivial non-invertible TDL $W$ has the following fusion rule and quantum dimension
\ie 
W^2=\id\oplus W\,,\quad \la W\ra=\zeta\equiv {\sqrt{5}+1\over 2}\,.
\fe 
The ${\rm \bf Fib}$ symmetry is realized in large class of CFTs including the $M(4,5)$ Virasoro minimal model and the $(G_2)_1$ Wess-Zumino-Witten CFT. The unique binary algebra in ${\rm \bf Fib}$ is
\ie 
A=\id \oplus W\,.
\fe
In the convention where (see e.g. \cite{Chang:2018iay})
\ie 
F^W_{WWW}(\id,\id)=F^W_{WWW}(\id,W)=-F^W_{WWW}(W,W)={1\over \zeta}\,,\quad F^W_{WWW}(W,\id)=1\,,
\fe
the multiplication morphism is fixed by \eqref{binary_qua_eqn} to
\ie 
m^{\id}_{\id\id}=m^{W}_{\id W}=m^{W}_{ W\id}=m^{\id}_{WW}=\zeta m^W_{WW}={1\over \zeta}\,.
\fe

We comment briefly on the case where the trivalent junction vector space for the self-dual simple TDL $\cL_i$ has dimension $N_{ii}^i > 1$ and discuss the constraint set upon the permutation map $F_{iii}^\id (\cL_i,\cL_i)$ by the quadratic equations \eqref{binary_qua_eqn}, which are implied by the existence of the binary algebra.\footnote{Note that the pentagon equations also constrain the permutation map of the trivalent junction vector space for $\cL_i$ \cite{Chang:2018iay}.} In general, the different components of the multiplication morphism for $\cL_i$ are related as follows,
\ie
m_{ii}^{i,\alpha} =  F_{iii}^\id (\cL_i,\cL_i)_{\beta\alpha} m_{ii}^{i,\beta} \,.
\label{qua_eqn_for_perm}
\fe
One can choose a basis of the trivalent junction vector space for $\cL_i$ such that $F_{iii}^\id (\cL_i,\cL_i)_{\beta\alpha}$ is diagonal, and~\eqref{qua_eqn_for_perm} can be rewritten as 
\ie 
m_{ii}^{i,\alpha} = \omega_{\alpha} m_{ii}^{i,\alpha}\,,\quad \alpha = 1, \dots, N_{ii}^i \,,
\fe
where each $\omega_{\alpha}$ is a third root of unity. For any $\omega_{\alpha} \neq 1$, we immediately obtain $m_{ii}^{i,\alpha}=0$. The first message conveyed by this constraint is that we must have at least one direction in the trivalent junction vector space which is invariant under the permutation map in order to have an algebra object (from \eqref{binary_qua_eqn}\footnote{Technically \eqref{binary_qua_eqn} is written for the case $N_{ii}^i = 1$, but a general form can be straightforwardly written down from it; we refer to its generalized form when commenting on the cases $N_{ii}^i > 1$. }). 
When $N_{ii}^i = 1$, this equation reduces to the constraint $F_{iii}^\id(\cL_i,\cL_i) = 1$, which means the permutation map has to be trivial for the binary algebra to exist. In addition, when $N_{ii}^i > 1$, if the dimension of the permutation invariant subspace in the trivalent junction vector space is 1, the quadratic equations \eqref{binary_qua_eqn} can be used to determine the multiplication morphism $m_{ii}^{i,\alpha}$ completely. An example is the $\frac{1}{2} E_6$ fusion category with simple TDLs $\{\id,X,Y\}$, where the trivalent junction vector space for $Y$ is two-dimensional and there is only one direction that is invariant under the permutation map so we can solve the multiplication morphism of the binary algebra object $A = \id \oplus Y$ from~\eqref{binary_qua_eqn} without solving the full system of pentagon equations,
\ie
m_{\id\id}^\id = m_{\id Y}^Y = m_{Y\id}^Y = m_{YY}^\id = \frac{1}{\sqrt{2 + \sqrt{3}}}\,, \;m_{YY}^Y = \frac{1}{\sqrt{2 + \sqrt{3}}} \left(\frac{1}{\sqrt{2}}\,, \frac{\sqrt{3} - 1}{2} \right) \,.
\fe
This is derived in the gauge where
\ie
F_{YYY}^\id (Y,Y) = \frac{1}{\sqrt{2}}\begin{pmatrix}
    e^{-\pi i/12} & e^{\pi i/6} \\
    e^{\pi i/6} & e^{-7\pi i/12}
\end{pmatrix}\,,\quad F_{YYY}^Y(Y,\id) = I \,.
\fe

\subsubsection{Gauging Binary Algebras in $SU(2)_{10}$ CFT}
\label{sec:SU210TDL}

For illustration, let us consider gauging binary algebras in the Wess-Zumino-Witten (WZW) CFT for $SU(2)$ symmetry. As an RCFT, it is constructed from $k+1$ unitary integrable representations of the affine Kac-Moody algebra $\widehat{\mf{su}(2)}_{k}$ for $k\in \mZ_{\geq 1}$, labeled by $0\leq j\leq {k\over 2}$ (see \cite{francesco2012conformal} for details). The RCFT is completely specified by a choice of modular invariant; this modular invariant then falls into the famous ADE classification, which includes the $A_n$-type or diagonal modular invariant at all $k=n-1$, the $D_n$-type or charge-conjugacy modular invariant at even $k=2n-4$ for $n\geq 4$, and the $E_6,E_7,E_8$ -type exceptional modular invariants at $k=10,16,28$, respectively \cite{Cappelli:1987xt,Kato:1987td}. In particular, the diagonal modular invariant RCFT is particular simple and commonly known as the $SU(2)_k$ WZW CFT. Its operator spectrum consists of $k+1$ scalar primary operators in one-to-one correspondence with the unitary integrable representations of $\widehat{\mf{su}(2)}_{k}$. 
Correspondingly there are $k+1$ Verlinde TDLs that we label as $\cL_j$, all of which are self-dual, and the Frobenius-Schur indicators are given by $\ep_j=(-1)^{2j}$. The fusion rules of the Verlinde TDLs are 
\ie 
\cL_i \cL_j =\cL_{|i-j|} +\cL_{|i-j|+1}+\dots +\cL_{\min(i+j,k-i-j)}\,,
\label{su2kverlindefusion}
\fe
which generate the fusion category 
\ie 
\cC(\mf{su}(2),k)\subset \cC_{SU(2)_k}\,,
\fe 
as a subcategory of the full symmetry category $\cC_{SU(2)_k}$ of the $SU(2)_k$ CFT.

In the modern point of view all other RCFTs with respect to $\widehat{\mf{su}(2)}_{k}$ are related to the $SU(2)_k$ WZW CFT by generalized discrete gaugings that commute with chiral algebra $\widehat{\mf{su}(2)}_{k}$ (i.e. with TDLs in $\cC(\mf{su}(2),k)$), and the different modular invariants correspond to inequivalent indecomposable module categories for $\cC(\mf{su}(2),k)$ \cite{Fuchs:2002cm}. Indeed, one recovers the ADE classification of \cite{Cappelli:1987xt,Kato:1987td} this way \cite{ostrik2001module,Fuchs:2002cm}. In particular, the algebra objects (up to Morita equivalence) relevant for the $D_n$-type and $E_6$-type modular invariants are binary algebras of the type \eqref{binaryalgebra}, which we will discuss in more detail below.

As an explicit example, consider the $SU(2)_{10}$ WZW CFT. By solving the quadratic equations in \eqref{binary_qua_eqn} with the explicit F-symbols for $\cC(\mf{su}(2),k)$), which descend from the $6j$-symbols for the  quantum group $U_q(\mf{sl}(2))$ \cite{kirillov1990representations} (we use the particular expressions in \cite{Fuchs:2002cm}), we find three binary algebra objects,
\ie 
A_{A_{11}}=\cL_0\oplus \cL_1\,,\quad A_{D_7}=\cL_0\oplus \cL_5\,,\quad A_{E_6}=\cL_0\oplus \cL_3\,,
\label{su2bA}
\fe
with unique multiplication morphisms that are in one-to-one correspondence with the three indecomposable module categories for $\cC(\mf{su}(2),10)$ and correspondingly the three possible modular invariants of $A_{11},D_{7},E_6$-types.

The first algebra $A_{A_{11}}$ in \eqref{su2bA} is Morita trivial in $\cC(\mf{su}(2),10)$ as $A_{A_{11}}=\cL_{1/2}\otimes \cL_{1/2}$ and thus corresponds to trivial gauging in the $SU(2)_{10}$ CFT. 
The second algebra $A_{D_7}$ in \eqref{su2bA} corresponds to gauging the non-anomalous $\mZ_2$ symmetry generated by the invertible TDL $\cL_5$ with trivial FS indicator. The multiplication morphism is simply,
\ie
m_{00}^0 = m_{05}^5 = m_{50}^5 = m_{55}^0 = \frac{1}{\sqrt{2}} \,.
\fe
Gauging $A_{D_7}$ produces the RCFT with D-type modular invariant, also known as the $SO(3)_5$ CFT. 

The more interesting case is the algebra $A_{E_6}$ in \eqref{su2bA}, which comes with the following multiplication morphism\footnote{Here we work with the F-symbols of \cite{Fuchs:2002cm}, which are in a gauge naturally inherited from the quantum group $6j$-symbols. However, it does not satisfy the positive gauge condition in Figure~\ref{fig:Fpositivegauge}, and this leads to the factor of $i$ in \eqref{E6m}. In particular, the comultiplication morphism differs from the adjoint of $m$ by $m^\vee=- m^\dagger$ in this case.}
\ie
m_{00}^0 = m_{03}^3 = m_{30}^3 = m_{33}^0 = \frac{1}{\sqrt{3 + \sqrt{3}}}\,, ~m_{33}^3 = \frac{2^{1/4} i}{\sqrt{3 + \sqrt{3}}} \,.
\label{E6m}
\fe
This algebra describes the non-invertible gauging from the $SU(2)_{10}$ CFT to the $Spin(5)_1$ CFT. In particular, in the gauged theory, there is an emergent larger chiral algebra $\widehat{\mf{so}(5)}_{1}$ (which contains the subalgebra $\widehat{\mf{su}(2)}_{10}$ via a conformal embedding), coming from extra conserved currents in the $\cL_3$-twisted sectors being liberated after gauging the binary algebra containing $\cL_3$.\footnote{In the $SU(2)_{10}$ CFT these conserved currents are point-like local operators attached to the $\cL_3$ TDL, and they produce a continuous family of TDLs for each $\cL_i$ such that $\cL_3\in \cL_i\otimes \cL_i$ following the general argument in \cite{Thorngren:2021yso}. Here we see such continuous families have a natural interpretation in the gauged theory via symmetry breaking. Namely $\mf{su}(2)$ preserving but $\mf{so}(5)$ breaking TDLs in the $Spin(5)_1$ CFT necessarily come in continuous families induced by a symmetry transformation in the bulk CFT.}

Since all generalized discrete gauging can be reversed, let us discuss the dual algebra objects in the gauged theory that implements the reverse gauging to the $SU(2)_{10}$ CFT. For the $SO(3)_5$ CFT, this amounts to gauging the charge conjugation symmetry, which is the quantum symmetry for the $\mZ_2$ generated by $\cL_5$ in the $SU(2)_{10}$ CFT. On the other hand, reversing the non-invertible gauging by $A_{E_6}$ in \eqref{su2bA}, requires a dual haploid algebra $B \in \cC_{Spin(5)_1}$ in the symmetry category of $Spin(5)_1$ CFT of quantum dimension (see \eqref{qdimAAdual}),
\ie 
\la B \ra = \la A_{E_6} \ra = 3+\sqrt{3}\,.
\label{Bdim}
\fe
The $Spin(5)_1$ CFT, in terms of the enhanced chiral algebra $\widehat{so(5)}_1$, contains three primary operators, and, correspondingly, three Verlinde TDLs which generate a fusion category ${\rm \bf Ising}$, with familar TDls $\{\id,\eta,\cN\}$. In addition, $\cC_{Spin(5)_1}$ contains two ${1\over 2}E_6$ fusion category symmetries \cite{hagge2007some,ostrik2013pivotal} related by conjugation (parity flip) \cite{Chang:2018iay}. However, it is easy to see all TDLs thereof have too large quantum dimensions to realize $B$ with \eqref{Bdim}. In \cite{DLWW2}, we determine the dual symmetry category $\cC_{\rm dual} \subset \cC_{Spin(5)_1}$ completely for gauging $A_{E_6} \in \cC(\mf{su}(2),10)$, which contains 12 simple TDLs. The $\cC_{\rm dual}$ category can be realized by the quotient of the following Deligne tensor product,
\ie 
\cC_{\rm dual}= \cC_{6} \boxtimes \cC_{6}^{\rm op} /\sim \,,
\label{CdualE6}
\fe
where $\cC_6$ is a rank 6 fusion category that comes from a $\mZ_2$-extension of the ${1\over 2}E_6$ category. Explicitly, the simple TDLs in $\cC_6$ are $\{\id,\cD_1,\cD_2,\cD_3,\eta,\cN\}$,  
which are all self-dual. The TDLs $\{\id,\eta,\cN\}$ generate the obvious ${\rm \bf Ising}$ subcategory that is identified in \eqref{CdualE6} among the two tensor factors.
The following fusion rules in $\cC_6$ are commutative and given by
\ie 
\cD_1^2=\cD_2^2=\id \oplus \cD_3\,,\quad \cD_3^2=\id \oplus \eta \oplus 2\cD_3\,,\quad \cD_1 \eta=\cD_2\,,\quad \cD_1 \cN=\cD_3\,, 
\label{C4fusion}
\fe 
where we have omitted those that follow from associativity. In particular the ${1\over 2} E_6$ subcategory is identified with the following subset of TDLs $\{\id,\cD_3,\eta\}$. 

From the constraint on the quantum dimension \eqref{Bdim} and general arguments in Section~\ref{sec:binaryintro}, we find that the dual algebra that reverses the $A_{E_6}$ gauging is given by the following unique binary algebra,
\ie 
B=\id \oplus \cD_1''\,,\quad \cD_1''\equiv \cD_1\cD'_1
\label{Balg}
\fe
where $\cD_1,\cD'_1$ come from the two $\cC_6$ factors in \eqref{CdualE6}, respectively, and have quantum dimensions
\ie 
\la \cD_1 \ra= \la \cD_1'\ra ={1+\sqrt{3} \over \sqrt{2}}\,.
\fe
It follows from \eqref{C4fusion} and \eqref{CdualE6} that $N_{\cD_1'' \cD_1''}^{\cD_1''} = 1$, and thus our analysis around \eqref{binary_qua_eqn} applies. Since the existence of this algebra is guaranteed by the general properties of gauging discussed in Section~\ref{sec:genprop}, the multiplication morphism can be fixed using \eqref{binary_qua_eqn} even without knowing the full set of F-symbols for $\cC_6$. We work in the conventional gauge where
\ie 
F_{\cD_1''\cD_1''\cD_1''}^{\cD_1''}(\cD_1'',\id) = 1\,,\quad 
F_{\cD_1''\cD_1''\cD_1''}^{\cD_1''}(\id,\id) = \la \cD_1''\ra = 2-\sqrt{3}\,,
\fe 
and we also know that $\cD_1''=\cD_1\cD_1'$ has the trivial FS indicator since $\cD_1$ and $\cD_1'$ are related by parity-flip and must have identical FS indicator.
The multiplication morphism for the dual binary algebra \eqref{Balg} is then given by\footnote{We thank Yu Nakayama for pointing out a typo in this formula. }
\ie
m_{\id\id}^{\id} = m_{\id\cD_1''}^{\cD_1''} = m_{\cD_1''\id}^{\cD_1''} = m_{\cD_1''\cD_1''}^{\id} = \frac{1}{\sqrt{3+\sqrt{3}}}\,,~m_{\cD_1''\cD_1''}^{\cD_1''} = \sqrt{\frac{\sqrt{3} - 1}{3 + \sqrt{3}}}\,.
\fe
It is straightforward to verify explicitly that the torus partition function of the $Spin(5)_1$ CFT decorated by $B$-mesh reproduces the diagonal modular invariant of the $SU(2)_{10}$ CFT.

 \section*{Acknowledgements}

We thank Chi-Ming Chang, Justin Kulp, Ying-Hsuan Lin, Victor Ostrik, Brandon Rayhaun, Sahand Seifnashri, Shu-Heng Shao, Noah Snyder, Constantin Teleman, and Ryan Thorngren for helpful questions and discussions.
OD and CL express gratitude to TASI 2023 summer school on ``Aspects of Symmetry"  for fostering an inspiring learning environment to explore generalized symmetries. The work of QW was partially supported by the NSERC PGS-D program. The work of YW was
supported in part by the NSF grant PHY-2210420 and by the Simons Junior Faculty Fellows program.

\bibliographystyle{JHEP}
\bibliography{refs}
 \end{document}